\documentclass[fleqn,usenatbib]{mnras}

\usepackage{newtxtext,newtxmath}

\usepackage[T1]{fontenc}
\usepackage{ae,aecompl}

\usepackage{graphicx}	
\usepackage{amsmath}	
\usepackage{amssymb}	
\graphicspath{{./figures/}}

\allowdisplaybreaks

\title[BH magnetosphere stability]{Black hole magnetosphere with small scale flux tubes--II. Stability and dynamics}

\author[Y. Yuan et al.]{
Yajie Yuan,$^{1}$\thanks{Lyman Spitzer, Jr. Postdoctoral Fellow.}\thanks{E-mail: yajiey@astro.princeton.edu (YY)}
Anatoly Spitkovsky,$^{1}$
Roger D. Blandford,$^{2}$
and Dan R. Wilkins$^{2}$\thanks{NASA Einstein Fellow.}
\\
$^{1}$Department of Astrophysical Sciences, Princeton University, Princeton, NJ 08544, USA\\
$^{2}$Kavli Institute for Particle Astrophysics and Cosmology (KIPAC), Stanford University, Stanford, CA 94305, USA
}

\date{Accepted XXX. Received YYY; in original form ZZZ}

\pubyear{2019}

\begin{document}
\label{firstpage}
\pagerange{\pageref{firstpage}--\pageref{lastpage}}
\maketitle

\begin{abstract}
In some Seyfert Galaxies, the hard X-rays that produce fluorescent emission lines are thought to be generated in a hot corona that is compact and located at only a few gravitational radii above the supermassive black hole. We consider the possibility that this X-ray source may be powered by small scale magnetic flux tubes attached to the accretion disk near the black hole. We use three dimensional, time-dependent,  special-relativistic, force-free simulations in a simplified setting to study the dynamics of such flux tubes as they get continuously twisted by the central compact star/black hole. We find that the dynamical evolution of the flux tubes connecting the central compact object and the accretion disk is strongly influenced by the confinement of the surrounding field. Although differential rotation between the central object and the disk tends to inflate the flux tubes, strong confinement from surrounding field quenches the formation of a jet-like outflow, as the inflated flux tube becomes kink unstable and dissipates most of the extracted rotational energy relatively close to the central object. Such a process may be able to heat up the plasma and produce strong X-ray emission. We estimate the energy dissipation rate and discuss its astrophysical implications.  
\end{abstract}




\section{Introduction}
In some Seyfert Galaxies there is a strong yet variable continuum X-ray source that irradiates the accretion disk and produces fluorescent emission lines. It is usually believed that the primary X-rays come from a hot corona located on either side of the accretion disk, where energetic electrons up-scatter optical/UV emission from the disk into X-ray range. Typical X-ray luminosity from the corona can be comparable to the disk thermal luminosity, indicating quite high X-ray efficiency. Recent observations have made good progress in constraining the geometry of the corona. Some pieces of evidence, including X-ray reverberation mapping \citep[e.g.,][]{Uttley2014A&ARv..22...72U,Kara2016MNRAS.462..511K}, disk emissivity profile modeling \citep{Wilkins2011MNRAS.414.1269W,Wilkins2015MNRAS.449..129W} and micro-lensing \citep{Morgan2008ApJ...689..755M,Chartas2009ApJ...693..174C,Mosquera2013ApJ...769...53M,ReisMiller2013ApJ...769L...7R}, suggest that the corona is quite compact and located at a few ($\lesssim20$) gravitational radii above the black hole, which seems consistent with the so-called ``lamppost'' geometry. However, the formation and heating mechanism of the corona is still unknown.

Another hint is that these Seyfert Galaxies usually do not create powerful jets. Meanwhile, their accretion rates are in the intermediate regime (the luminosity $L$ is in the range $0.01L_{\rm Edd}\lesssim L\lesssim L_{\rm Edd}$), so the accretion disk is likely to be geometrically thin \citep[e.g.,][]{2014ARA&A..52..529Y}. Given these circumstances, we consider the possibility that the X-ray corona may be powered by small scale magnetic flux tubes on the accretion disk around the black hole \citep[henceforth Paper I]{2019MNRAS.484.4920Y}. These flux tubes may have foot points located at different radii on the disk or one foot point on the disk and the other on the black hole. This kind of field configurations may arise if the large scale net flux is weak so that the inner disk is dominated by small coherence length scale structures produced by magneto-rotational instability (MRI) in the disk; or, similar to the solar surface, flux tubes may emerge from hot spots on the disk due to magnetic buoyancy. Differential rotation causes the flux tubes to twist and inflate \citep[e.g.,][]{1996A&A...313.1008A,1996A&A...313.1016A,UzdenskyGoodman2008ApJ...682..608U}; in particular, they may get tangled up on or near the axis. This may lead to strong dissipation as the field relaxes/untangles, and most of the extracted rotational energy is converted into particle kinetic energy, then radiated away as X-rays. In such a way, the black hole may directly power the X-ray corona, instead of launching a powerful jet.

In Paper I, we set up simplified models where the magnetic field attached to the disk and the black hole is still axisymmetric, but has relatively small coherence length scales along the radial direction---on the order of a few gravitational radii \citep[similar to][]{Parfrey2015MNRAS.446L..61P,Uzdensky2005ApJ...620..889U}. We found that, under the axisymmetric constraint, there exist steady configurations with disk-hole linking ``closed'' flux tubes, consistent with \citet{Uzdensky2005ApJ...620..889U}. Furthermore, the extent of the closed zone is determined by the pressure balance between the spin induced twist in the closed zone and the poloidal confinement from both the closed zone itself and the overlying external field from the disk. The maximum extent decreases with black hole spin; for a typical confinement and black hole spin, the closed zone does not exceed a height of tens of gravitational radii. In a real dynamic situation, the equilibrium may be unstable, or the pressure balance may not be easily maintained. The flux tubes may become highly dynamic and produce significant dissipation. It is the goal of this paper to follow the time-dependent evolution of the flux tubes. 

The paper is organized as follows. We describe our simplified setup and numerical method in \S\ref{sec:method}. The simulation results are presented in \S\ref{sec:result} and \S\ref{sec:dissipation}. We then discuss the implications and conclusions in \S\ref{sec:dicussion_conclusion}.

\section{Numerical approach}\label{sec:method}
\subsection{The Membrane paradigm}\label{subsec:membrane}
The boundary conditions for the electromagnetic fields at the event horizon of a Kerr black hole can be interpreted in a way such that the event horizon behaves like a resistive membrane \citep[e.g.,][]{thorne_black_1986}.
In particular, define the tangential magnetic field at the horizon as $\mathbf{B}_H=\alpha_H\mathbf{B}_{\parallel}$, where $\mathbf{B}_{\parallel}$ is the tangential magnetic field as measured by a fiducial observer (FIDO, taken to be ZAMO) close to the event horizon, and $\alpha$ is the lapse function there (though the magnitude of $\mathbf{B}_{\parallel}$ diverges at the event horizon, $|\alpha_{H}\mathbf{B}_{\parallel}|$ remains finite), then one can define a surface current $\pmb{\mathcal{J}}_H$ from Ampere's law by requiring that the tangential magnetic field terminates at the event horizon:
\begin{equation}
\mathbf{B}_H\equiv 4\pi \pmb{\mathcal{J}}_H\times\mathbf{n},
\end{equation}
where $\mathbf{n}$ is a unit vector normal to the event horizon.
The surface current is driven by the tangential component of the horizon's renormalized electric field and satisfies an Ohm's law
\begin{equation}
\mathbf{E}_H\equiv\alpha_H\mathbf{E}_{\parallel}=R_H\pmb{\mathcal{J}}_H,
\end{equation}
where the surface resistivity 
\begin{equation}
R_H\equiv 4\pi\approx 377\,\text{ohms}.
\end{equation}
The Ohm's law and Ampere's law are equivalent to the radiative boundary condition at the event horizon
\begin{equation}
\mathbf{B}_H=\mathbf{E}_H\times\mathbf{n}.
\end{equation}
If one also defines the surface charge density $\sigma_H$ by Gauss's law,
\begin{equation}
\sigma_H\equiv E_n/(4\pi),
\end{equation}
where $E_n$ is the normal component of the electric field as measured by a FIDO at the event horizon, then one can obtain the charge conservation relation
\begin{equation}
\partial\sigma_H/\partial t+^{(2)}\nabla\cdot\pmb{\mathcal{J}}_H+(\alpha_Hj_n)=0,
\end{equation}
where $^{(2)}\nabla\cdot\pmb{\mathcal{J}}_H$ is the 2-dimensional divergence of the 2-dimensional surface current; $j_n$ is the normal component of the volume current leaving the horizon as measured by a FIDO.

In our flat spacetime toy model, we take into account the electromagnetic property of the event horizon using a similar membrane formulation. Suppose the event horizon can be described as a resistive membrane, rotating at an angular velocity $\pmb{\Omega}$. A FIDO that is sitting on the membrane rotates with the membrane at the same velocity $\mathbf{v}=\pmb{\Omega}\times\mathbf{r}$ and sees a tangential magnetic field $\mathbf{B}_{\parallel}'$ terminating at the membrane (we use prime to denote quantities measured by the rotating FIDO), which requires a surface current density 
\begin{equation}
\pmb{\mathcal{J}}'=\mathbf{n}\times\mathbf{B}_{\parallel}'/(4\pi).
\end{equation}
The surface current needs to be driven by a tangential electric field
\begin{equation}
\mathbf{E}_{\parallel}'=R \mathbf{\mathcal{J}}',
\end{equation}
where $R$ is the surface resistivity. To mimic a black hole we take $R=4\pi$. For a general Ohmic resistivity $R$, the above relations give (in what follows we write $R/4\pi\equiv\eta$)
\begin{equation}\label{eq:membrane-boundary}
\mathbf{B}_{\parallel}'=\frac{1}{\eta}\mathbf{E}_{\parallel}'\times\mathbf{n}.
\end{equation}
The FIDO-measured fields are related to the lab frame values through a Lorentz transformation
\begin{align}
\mathbf{E}_{\parallel}'=\gamma(\mathbf{E}_{\parallel}+\mathbf{v}\times\mathbf{B}_{\perp})-(\gamma-1)(\mathbf{E}_{\parallel}\cdot\hat{v})\hat{v},\label{eq:E_Lorentz}\\
\mathbf{B}_{\parallel}'=\gamma(\mathbf{B}_{\parallel}-\mathbf{v}\times\mathbf{E}_{\perp})-(\gamma-1)(\mathbf{B}_{\parallel}\cdot\hat{v})\hat{v},\label{eq:B_Lorentz}
\end{align}
where $\gamma=1/\sqrt{1-v^2/c^2}$ is the Lorentz factor of the FIDO.

In Appendix \ref{sec:membrane_test} we show a comparison between steady state configurations obtained using the above membrane boundary conditions in flat spacetime and those for exact Kerr spacetime. The similarity lends further support that the membrane analogy captures the electromagnetic properties of the black hole reasonably well.

\subsection{Simulation setup}
We consider the force-free limit of magnetohydrodynamics, which is appropriate when the magnetization is sufficiently high. In flat spacetime, the system of equations can be written as \citep[e.g.,][]{1999astro.ph..2288G,2002luml.conf..381B}
\begin{align}
\frac{\partial\pmb{E}}{\partial t}&=c \nabla\times\pmb{B}-4\pi \pmb{J},\\
\frac{\partial\pmb{B}}{\partial t}&=-c \nabla\times\pmb{E},\\
\pmb{J}&=\frac{c}{4\pi}\nabla\cdot\pmb{E}\frac{\pmb{E}\times\pmb{B}}{B^2}+\frac{c}{4\pi}\frac{(\pmb{B}\cdot\nabla\times\pmb{B}-\pmb{E}\cdot\nabla\times\pmb{E})\pmb{B}}{B^2}.
\end{align}
We use the three dimensional, Cartesian, special relativistic force-free code developed by \citet{2006ApJ...648L..51S} to follow the evolution of the system. To mimic the black hole, we put a disk-like membrane with radius $r_1$ and angular velocity $\Omega$ at the center of the $z=0$ plane, on which we apply the boundary condition described in \S\ref{subsec:membrane}. More specifically, we specify $\pmb{E}_{\parallel}$ using equations (\ref{eq:membrane-boundary}--\ref{eq:B_Lorentz}): 
\begin{align}
E_r&=-\eta\frac{B_{\phi}}{\gamma}-\frac{\Omega r}{c}B_z,\label{eq:bc_Er}\\
E_{\phi}&=\eta\gamma(B_r-\frac{\Omega r}{c}E_z).\label{eq:bc_Ephi}
\end{align}
Here $r$ denotes the radial direction on the disk-like membrane, and $\phi$ indicates the azimuthal direction.
The ``black hole'' (or rather, ``black disk'') is surrounded by a thin accretion disk, modeled by a perfectly conducting surface extending over the region $r>r_2$ on the $z=0$ plane. The inner boundary of the disk $r_2$ may or may not correspond to the innermost stable circular orbit (ISCO). In what follows we show two classes of examples: one has $r_2\gtrapprox r_1$ to mimic a rapidly spinning black hole with a disk extending close to the event horizon; the other has $r_2=2r_1$ to represent a disk further away from the hole (e.g., truncated disk). In both cases the angular velocity of the central compact object satisfies $\Omega r_1/c=0.9$. We have experimented with different $\Omega$ values and found the results qualitatively similar. 
Suppose the angular velocity of the disk is $\Omega_d(r)$, then the boundary condition on the accretion disk is
\begin{equation}
E_r=-\frac{\Omega_d r}{c}B_z.
\end{equation}
To minimize the effect of Cartesian grid stair-stepping at the boundary of the disk and the membrane, we apply smoothing of the fields over a few grid cells\footnote{Near the edge of the membrane, we apply the boundary condition as $E_{\parallel}=s E_{\parallel,\rm membrane}+(1-s)E_{\parallel,\rm local}$, where $s=[1-\tanh((r-r_1)/\delta)]/2$ and the smoothing length scale $\delta$ is 1 to 2 grid cells.}. For the outer boundary condition of the simulation domain, we use an absorbing layer that damps all electromagnetic disturbances. The absorbing layer satisfies the modified Maxwell equations that contain an electric and magnetic conductivity term \citep[e.g.,][]{2015MNRAS.448..606C} 
\begin{align}
\frac{\partial \pmb{E}}{\partial t}&=-\lambda(\pmb{r})\pmb{E}+c\nabla\times\pmb{B}-4\pi\pmb{J},\\
\frac{\partial \pmb{B}}{\partial t}&=-\lambda^*(\pmb{r})\pmb{B}-c\nabla\times\pmb{E},
\end{align}
where the conductivity varies with distance $d$ to the boundary of the physical domain
\begin{equation}
\lambda(\pmb{r})=\lambda^*(\pmb{r})=\frac{K_{\rm abs}}{\Delta t}\left(\frac{d}{d_{\rm max}}\right)^3.
\end{equation}

\begin{figure*}
\includegraphics[width=\textwidth]{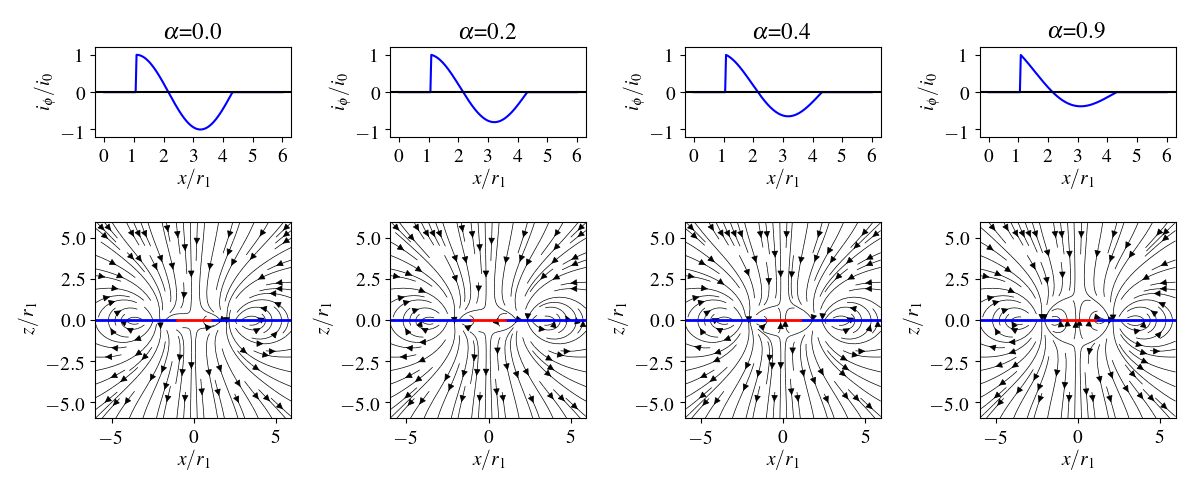}
\caption{A few examples of the initial magnetic field configurations. These are potential fields generated by toroidal current distributions $i_{\phi}$ in the disk specified by Equation (\ref{eq:iphi}). The top row shows the profile of $i_{\phi}$ on the disk and the bottom row shows the corresponding magnetic field lines on the meridional plane. We vary the exponent $\alpha$ to change the relative strength between the two oppositely directed current components. Here we fix the radius of the inner boundary of the disk $r_2$ to be slightly larger than the radius of the membrane $r_1$, such that $r_2=1.08r_1$. In the bottom row, the red line represents the membrane and the blue line indicates the accretion disk.}\label{fig:initial}
\end{figure*}

We start with a vacuum potential magnetic field generated by a surface current distribution $i_{\phi}$ in the disk, then turn on the rotation of the ``black hole'' and the accretion disk, and watch how the field evolves. In what follows, for simplicity, we set the accretion disk to be non-rotating: $\Omega_d(r)=0$, which is a rough approximation when the central object is rotating with a rate $\Omega$ much larger than every radius of the disk.\footnote{For a real black hole, if the inner boundary of the disk is at the ISCO, then the black hole rotates faster than every radius of the disk when its spin $a>0.3594$, or equivalently, the angular velocity of the event horizon $\Omega_H>0.093c/r_g$.}
We will also briefly discuss the effects of the disk rotation in section \ref{subsec:disk_rotation}.

The initial disk current distribution is
\begin{equation}\label{eq:iphi}
i_{\phi}=
\begin{cases}
\displaystyle i_0\cos\left(\frac{2\pi}{r_0}(r-r_2)\right)\left(\frac{r_2}{r}\right)^{\alpha}, & \displaystyle r_2\le r\le r_2+\frac{3}{4}r_0,\\
\displaystyle 0, & \text{otherwise}.
\end{cases}
\end{equation}
The current distribution has two oppositely directed components (see the first row of Figure \ref{fig:initial}); $r_0$ and the exponent $\alpha$ control the separation and relative strength of the two components, respectively. The resulting field geometry has two dipole-like loops (the second row of Figure \ref{fig:initial}). As $\alpha$ increases, the outer current loop becomes weaker, so does the outer magnetic field. As a result, the inner magnetic field loop becomes larger in size. 

In our toy model the magnetic flux is fixed in the accretion disk; this could be a reasonable assumption if the overall flux changes on the accretion time scale, which is much longer than the orbital time scale relevant for our simulation here.

\section{Simulation results: the dynamics}\label{sec:result}
\subsection{The case of $\eta=0$: a perfectly conducting star}
\begin{figure*}
\includegraphics[width=\textwidth]{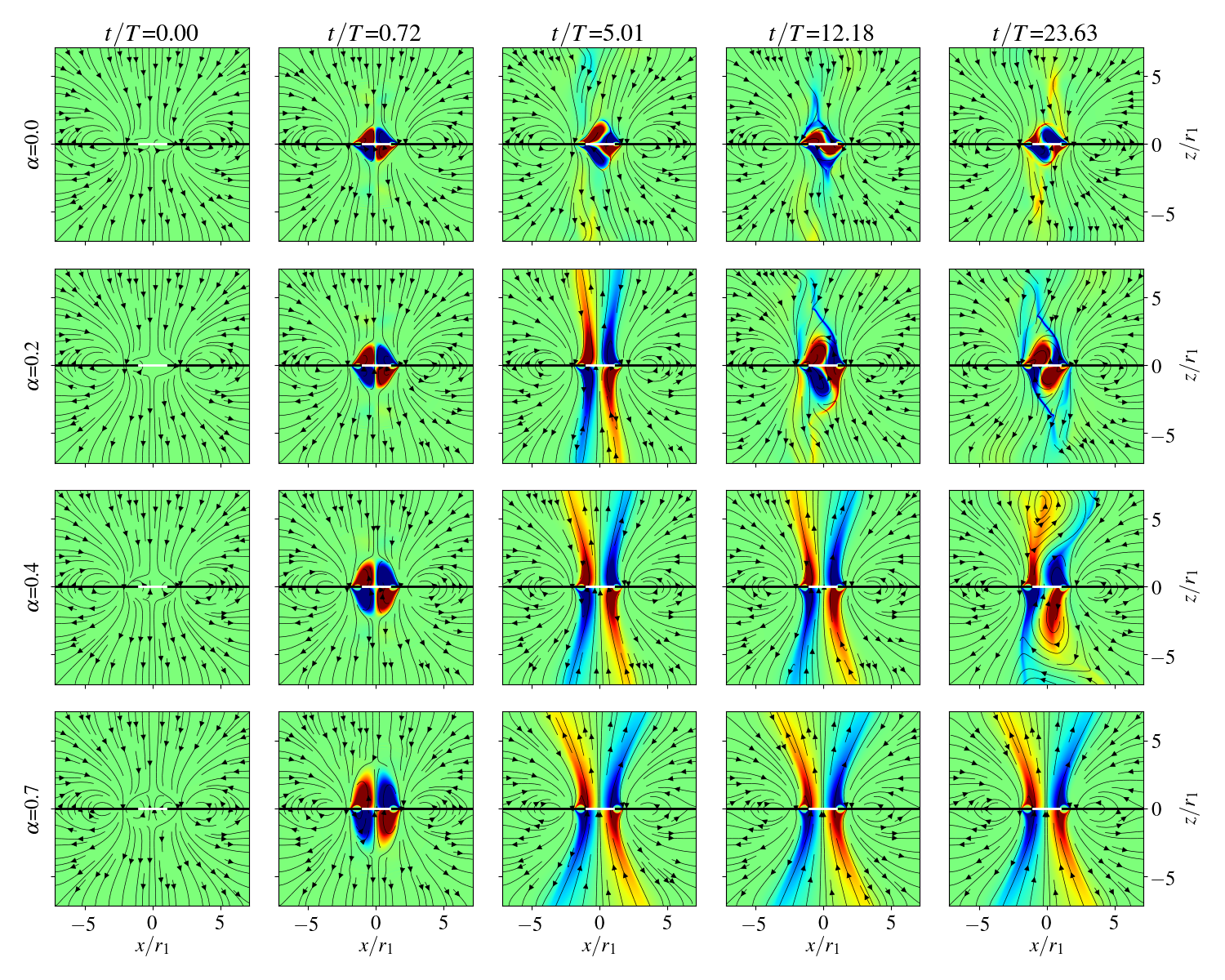}
\caption{Snapshots from simulations where the central membrane has zero resistivity. The four rows show four sets of simulations with different $\alpha$ such that the relative strength of the two current loops on the disk changes. All simulations have $r_2=1.08r_1$, and the membrane angular velocity satisfies $\Omega r_1/c=0.9$. From top to bottom: $\alpha=0.0,0.2,0.4,0.7$. The thick white line on the equatorial plane marks the membrane and the thick black lines on the equatorial plane represent the (non-rotating) accretion disk. Stream lines show magnetic field components on the $x-z$ plane, while colors show the component perpendicular to the plane ($B_y$, red is positive and blue is negative). All panels have the same color scale. Time is in units of the rotational period of the membrane $T=2\pi/\Omega$ (same below).}\label{fig:eta0slice}
\end{figure*}

\begin{figure*}
\includegraphics[width=\textwidth]{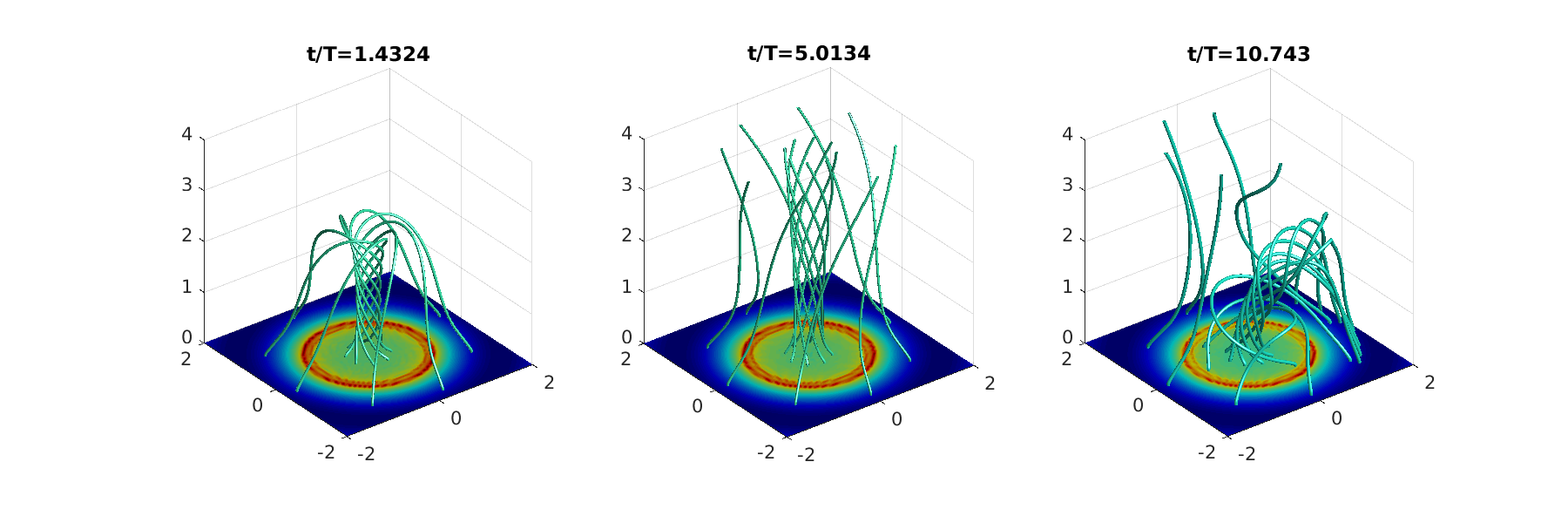}
\caption{3D rendering of field line configurations at a series of time points in a simulation where the central membrane has $\eta=0$ and the current in the disk has $\alpha=0.2$ (corresponding to the second row of Figure \ref{fig:eta0slice}). Color on the equatorial plane shows $B_z$. Left panel: buildup of the magnetic tower; middle panel: open-up of the magnetic field; right panel: fully developed kink. The lengths are scaled to the radius of the membrane.}\label{fig:alpha0.2eta0-3d}
\end{figure*}

When we set $\eta=0$, equation (\ref{eq:bc_Er}) and (\ref{eq:bc_Ephi}) indicate that the central membrane behaves like a perfectly conducting star. Any field line that threads the star will have its foot point corotating with the star. If the other end is attached to the disk, differential rotation between the disk and the star will build up more and more twist along the field line. Normally under axisymmetric constraint, the magnetic field will be inflated by the increasing toroidal twist, and eventually open up to infinity \citep[e.g.,][]{2002ApJ...565.1191U}. This is a direct consequence of the relativistic Ferraro's Law of isorotation \citep[e.g.,][]{Blandford1977MNRAS.179..433B}: if a steady state can be reached in axisymmetry, the angular velocity along the field line must be a constant. However, in a real 3D case, the situation can be different.
\subsubsection{Different behaviors as $\alpha$ changes}
In the first series of experiments, we fix the inner boundary of the `accretion disk' at $r_2$ and the current loop separation at $r_0$, but change the exponent $\alpha$ to vary the relative strength of the two current loops. We find qualitatively different behaviors when $\alpha$ changes.

For sufficiently large $\alpha$, the outer current loop is relatively weak compared to the inner one. As we turn on the rotation of the star, a torsional Alfv\'{e}n wave is emitted from the star and propagates out along the field lines threading the star. However, the other end of the field line is attached to the disk and is nonrotating. Therefore these field lines get twisted up, building up a magnetic tower \citep[e.g.,][]{1996MNRAS.279..389L} which eventually breaks out from the overlying field. The inner field loop becomes open as a result, and there is a current layer separating the two polarities of magnetic field (Figure \ref{fig:eta0slice}). This current layer marks the `jet wall': within the jet wall, the field is rotating with the angular velocity of the `star' and there is a Poynting flux going out to infinity; outside the jet wall, the field is attached to the accretion disk and is non-rotating.

When we decrease $\alpha$, the opening angle of the `jet' decreases, as can be seen clearly in the middle column of Figure \ref{fig:eta0slice}. For small enough opening angle, we start to see the development of kink instability in the jet. The kink mode growth rate increases as the opening angle decreases. For very small $\alpha$, namely, when the outer current loop is much stronger than the inner one, we do not see the jet break out: the kink mode develops very early and completely destroys the outflow. 
Figure \ref{fig:alpha0.2eta0-3d} shows 3D renderings of field lines at three different time slices for the simulation with $\alpha=0.2$, where the inner flux loop first opens up, then gets destroyed by the kink instability. When the kink is fully developed, we see that the field lines from the star connect to one side of the accretion disk, forming a closed structure. As the star rotates, the closed structure precesses---its foot points on the disk appear to rotate around the star, in an opposite direction, with an apparent angular velocity that is about half of the angular velocity of the star. The field lines in and around this closed structure are actually continuously changing partners, or equivalently, reconnecting: if a particular point on the star and a particular point on the disk is connected by a field line at some point, they can only maintain the connectivity for a short amount of time; the rotation of the star will increase the toroidal twist in this connecting field line, which eventually breaks---the disk-end becomes an open field line, while the star-end gets connected to an open field line from the opposite side of the disk. The star-end remains closed, but it is now connected to a different point on the disk. As a result, the field lines on the accretion disk around the closed zone alternate between closed and open configurations, with continuous reconnection happening at the separatrix layer that lies between the open and closed field lines. A movie corresponding to this simulation is available \href{https://youtu.be/htSQFD64JGI}{online}\footnote{Direct link: \url{https://youtu.be/htSQFD64JGI}}.

\subsubsection{The criterion for kink instability}
\begin{figure*}
\includegraphics[width=\textwidth]{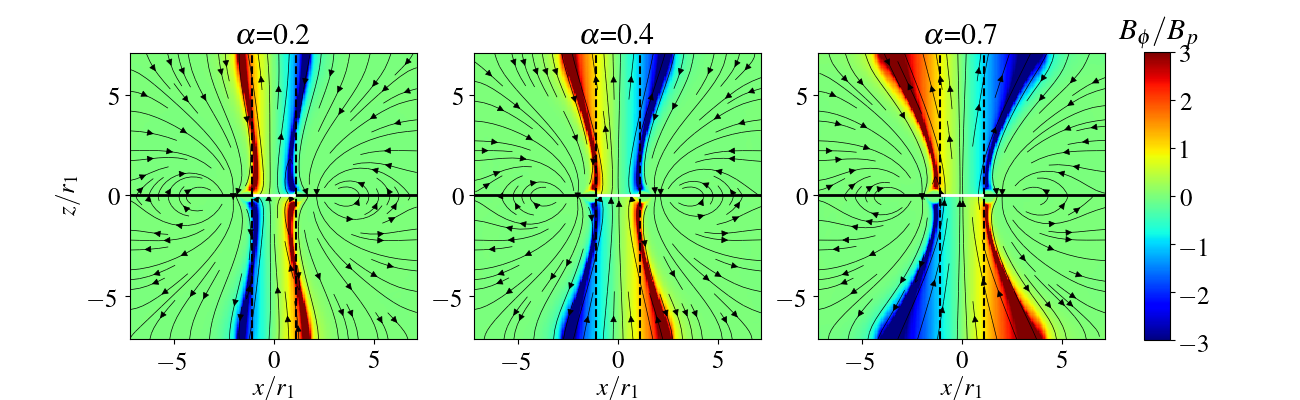}
\caption{The profile of $B_{\phi}/B_p$ on the meridional plane for the three cases $\alpha=0.2,0.4,0.7$, at $t=5.01T$, corresponding to the middle column of Figure \ref{fig:eta0slice}. The vertical dashed lines correspond to the light cylinders.}\label{fig:BphioBp}
\end{figure*}

\begin{figure*}
\includegraphics[width=\textwidth]{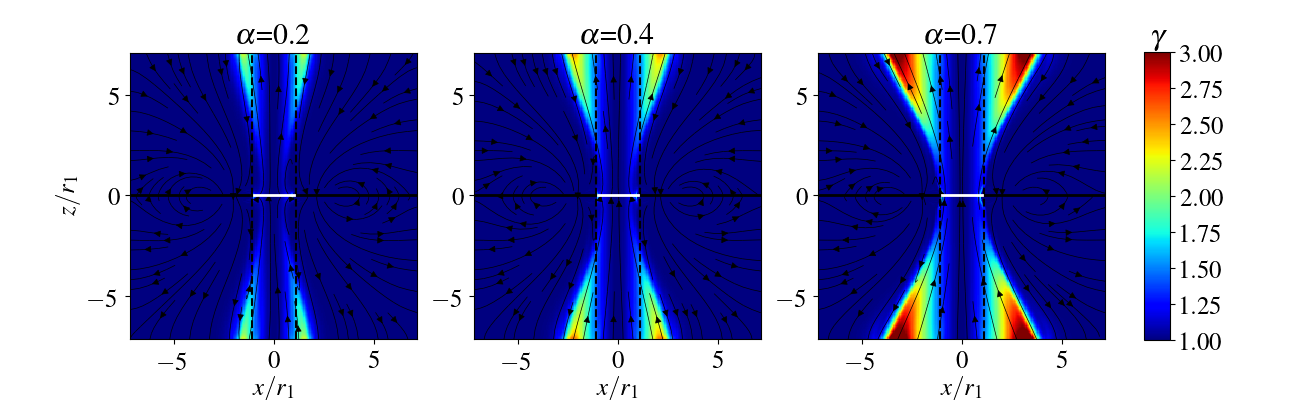}
\caption{Lorentz factor of the outflow for the three cases $\alpha=0.2,0.4,0.7$, at $t=5.01T$, corresponding to the middle column of Figure \ref{fig:eta0slice}. The vertical dashed lines correspond to the light cylinders.}\label{fig:eta0gm}
\end{figure*}

\begin{figure}
\includegraphics[width=\columnwidth]{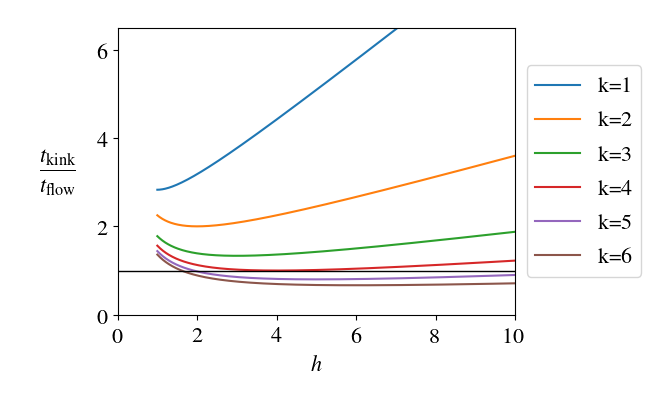}
\caption{The ratio $t_{\rm kink}/t_{\rm flow}$, as a function of the dimensionless height $h=z/r_{\rm LC}$, for different opening angle profile $z=k(r-r_{\rm LC})$. The black horizontal line corresponds to a ratio of 1.}\label{fig:eta0kink}
\end{figure}

In these configurations, when the field lines attached to the central membrane open up, the toroidal field becomes dominant toward the edge of the jet funnel (on the boundary the poloidal field reverses and goes through zero but the toroidal field is nonzero), as shown in Figure \ref{fig:BphioBp}. Jet-like outflows dominated by toroidal field are prone to kink instability, which grows on Alfv\'{e}n wave crossing time scales in the rest frame of the flow \citep[e.g.][]{1992SvAL...18..356L,2000A&A...355..818A,2015MNRAS.452.1089P,2016MNRAS.456.1739B}. If in the lab frame the kink growth time scale is shorter than the flow time scale, the jet can become kink unstable. Equivalently speaking, when the jet material stays in causal contact on the flow time scale, kink will be able to develop. In the above examples, we see that for different $\alpha$ the jet opening angle under axisymmetry constraint is different, and this affects the growth of the kink mode. We can obtain a rough criterion for the kink instability as follows.

The kink growth time scale in the rest frame of the flow is approximately 
\begin{equation}
t_{\rm kink}'=\frac{r}{v_A},
\end{equation}
where $r$ is the radius of the jet and $v_A$ is the Alfv\'{e}n velocity. In the lab frame, we have
\begin{equation}
t_{\rm kink}=\gamma\frac{r}{v_A},
\end{equation}
where $\gamma$ is the Lorentz factor of the flow.
From the simulation results we see that the jet boundary can be described approximately as
\begin{equation}
z=k(r-r_{\rm LC}),
\end{equation}
where $r_{\rm LC}$ is the radius of the light cylinder of the star, and $k$ is a constant characterizing the opening angle of the jet. The Lorentz factor of the flow can be estimated as \citep[e.g.,][]{1992SvAL...18..356L}
\begin{equation}
\gamma\approx r/r_{\rm LC}.
\end{equation}
This is exact for Michel monopole solution \citep{1973ApJ...180L.133M}, and is also a good approximation for our case, as can be seen in Figure \ref{fig:eta0gm}. Now the kink growth time scale is
\begin{equation}
t_{\rm kink}\approx\frac{r^2}{r_{\rm LC}v_A}\approx\frac{r^2}{r_{\rm LC}c}.
\end{equation}
On the other hand, the flow time scale is 
\begin{equation}
t_{\rm flow}\approx\frac{z}{v \cos\theta}\approx\frac{\sqrt{z^2+(r-r_{\rm LC})^2}}{c}.
\end{equation}
Writing $z=hr_{\rm LC}$, we get the ratio
\begin{equation}
\frac{t_{\rm kink}}{t_{\rm flow}}\approx\frac{(h/k+1)^2}{h\sqrt{1+1/k^2}}.
\end{equation}
Figure \ref{fig:eta0kink} shows the ratio $t_{\rm kink}/t_{\rm flow}$ as a function of $h$ for different opening angles. When $k$ gets larger than 4, namely the opening angle becomes smaller than $\sim 14^{\circ}$, $t_{\rm kink}/t_{\rm flow}$ can become less than 1 and the kink can develop. Our $\alpha=0.4$ case is close to marginal instability and agrees well with the above criterion.

\subsection{Effect of changing the membrane resistivity $\eta$}
\begin{figure*}
\includegraphics[width=\textwidth]{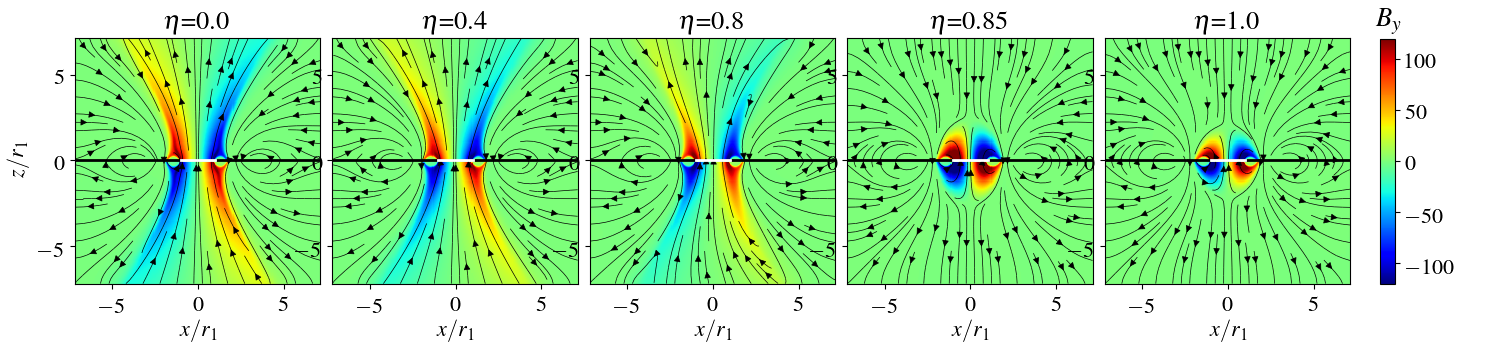}\\
\includegraphics[width=\textwidth]{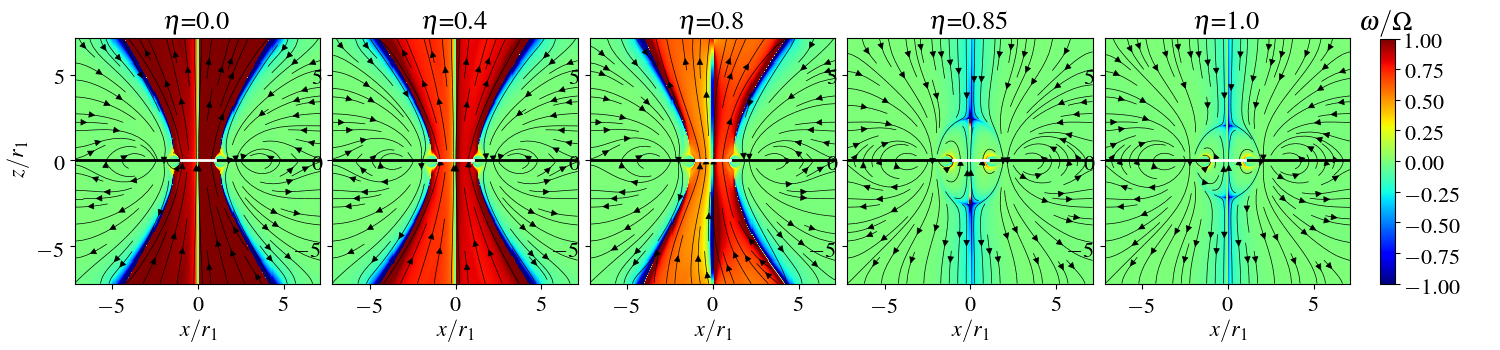}
\caption{A slice on the $x-z$ plane at time $t=23.63T$ from a series of simulations where the central membrane has different resistivity $\eta\equiv R/4\pi$. The current distribution in the disk is kept the same, with $r_2=1.08r_1$ and $\alpha=0.9$. Streamlines show the magnetic field component in the $x-z$ plane. In the top row, the color shows the component of magnetic field that is perpendicular to the plane ($B_y$), while in the bottom row, the color shows the angular velocity of the field line (valid in axisymmetric case, see Equation \ref{eq:omega}).  For small $\eta$ and large $\eta$, we do reach a steady state where the configuration remains axisymmetric. Around the transition between the open and closed states (e.g. $\eta=0.8$), we see kink instability developing.}\label{fig:alpha_0.9slice}
\end{figure*}

\begin{figure*}
\includegraphics[width=\textwidth]{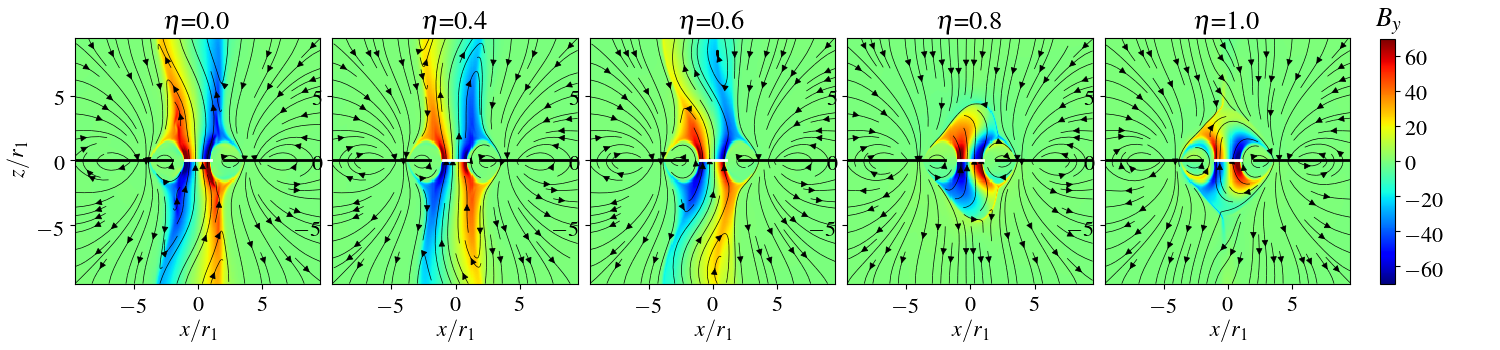}
\caption{A slice on the $x-z$ plane at time $t=22.92T$ from a series of simulations where the central membrane has different resistivity $\eta\equiv R/4\pi$. The current distribution in the disk is kept the same, with $r_2=2r_1$ and $\alpha=1.0$. The membrane angular velocity satisfies $\Omega r_1/c=0.9$. Streamlines show the magnetic field component in the $x-z$ plane, while the color shows the component of magnetic field that is perpendicular to the plane ($B_y$).}\label{fig:rin2_100slice}
\end{figure*}

Changing the membrane resistivity $\eta$ can also change the behavior of the system qualitatively.
In Figure \ref{fig:alpha_0.9slice}, we show a series of experiments where the current distribution in the disk is kept the same, while the membrane resistivity varies. We still set $r_2=1.08r_1$, and the exponent $\alpha$ is 0.9. For zero resistivity case, the inner field loop quickly opens up, and remains a stable steady state thereafter. The configuration is axisymmetric, and we can directly calculate the angular velocity $\omega$ of the field line in the following way: since 
\begin{equation}
\pmb{E}=-(\pmb{\omega}\times\pmb{r})\times\pmb{B}/c,
\end{equation}
the $\phi$ component of $\pmb{E}\times\pmb{B}$ is thus
\begin{equation}
(\pmb{E}\times\pmb{B})_{\phi}=\frac{\omega r\sin\theta}{c}B_p^2,
\end{equation}
where the subscript p denotes poloidal component. So we have 
\begin{equation}\label{eq:omega}
\omega=\frac{(\pmb{E}\times\pmb{B})_{\phi}}{B_p^2r \sin\theta}.
\end{equation}
The lower panels of Figure \ref{fig:alpha_0.9slice} show the angular velocity $\omega$ of the field line. Indeed, when $\eta=0$, $\omega=\Omega$.
As we increase the resistivity of the membrane, the angular velocity of the field line decreases. This is because the field lines can now slip on the membrane, and the resistivity allows them to rotate at a slower rate than the membrane itself. The toroidal field also decreases, meaning that the field lines are not twisted up as efficiently. 

Close to the center of our disk-like membrane, the magnetic field is nearly uniform and points along z direction. We can use the membrane boundary condition to determine the angular velocity of these field lines. Equation (\ref{eq:bc_Er}) suggests that on the membrane, we have
\begin{equation}
E_r=-\eta\frac{B_{\phi}}{\gamma}-\frac{\Omega r}{c}B_z.
\end{equation}
Right outside the membrane, the plasma is perfectly conducting so we have
\begin{equation}
E_r=-\frac{\omega r}{c}B_z,
\end{equation}
and the outgoing Alfv\'{e}n wave along the vertical uniform field has 
\begin{equation}
E_r=B_{\phi}.
\end{equation}
Using the three equations, we obtain the angular velocity of the field line to be
\begin{equation}\label{eq:omega(eta)}
\omega=\frac{\Omega}{1+\eta}
\end{equation}
when $\Omega r\ll c$, $\gamma\approx1$. In the simulations as we increase the resistivity $\eta$, the field line angular velocity decreases, agreeing with this analytical expression.

For sufficiently large resistivity, the inner magnetic flux loop no longer opens up. At $t=0$ when we turn on the rotation of the membrane, the membrane boundary condition still results in a torsional Alfv\'{e}n wave being emitted from the foot points of the field lines on the black hole. The Alfv\'{e}n wave propagates along the flux tubes, reaches the accretion disk, and gets reflected. The interaction between the outgoing wave and the reflected wave tries to establish an equilibrium configuration where the field lines linking the membrane and the accretion disk all have the same angular velocity as their foot points on the disk, namely, $\omega=\Omega_d=0$. Despite their zero angular velocity, the field lines do develop a toroidal component as they are dragged by the rotation of the membrane and have to slide on the membrane. From Figure \ref{fig:alpha_0.9slice} we can see that the vertical extent of the membrane-disk link decreases as the resistivity increases, because there is less twist in the magnetic field, or equivalently, less toroidal pressure pushing on the external field, when resistivity increases. This asymptotes to a vacuum potential field when we set $\eta$ to be very large. The existence and behavior of the closed zone with disk-hole linking magnetic flux agree with our results in Paper I.

We can also view this from a slightly different angle. The closed field lines lie in the sub-Alfv\'{e}nic regime, as such an equilibrium can only be established when signals---Alfv\'{e}n waves---can propagate back and forth along the field lines. Indeed, the initial Alfv\'{e}n pulse has to be reflected at some point (e.g., the foot point on the disk) in cases where a closed equilibrium develops. We find from our simulations that the necessary condition for this is that the reflection point lies within the effective light surface as determined by the angular velocity from Equation (\ref{eq:omega(eta)}) if the field lines were to open up. In Figure \ref{fig:alpha_0.9slice}, as the resistivity $\eta$ increases, the effective light surface is moved to larger radii, and when it is further away than the reflection point of the closed flux bundle, the field lines no longer open up.

Around the transition between the open final state and the closed final state, the system is prone to the kink instability, as shown by the middle column in Figure \ref{fig:alpha_0.9slice}. Although the opening angle of the outflow does not seem to change with the resistivity $\eta$, increasing $\eta$ does reduce the outflow velocity and puts the light surface at larger radius. When a large portion of the jet base lies within the light surfaces, namely within causal contact, kink instability will start to develop. Figure \ref{fig:rin2_100slice} shows another set of examples where we put the disk current further away from the central membrane, with $r_2=2r_1$ and $\alpha=1.0$. The configuration is prone to kink in all different resistivity cases, but the kink grows fastest near the open/closed transition.   

\subsection{The case of a ``black hole'' ($\eta=1$)}
\begin{figure*}
\includegraphics[width=\textwidth]{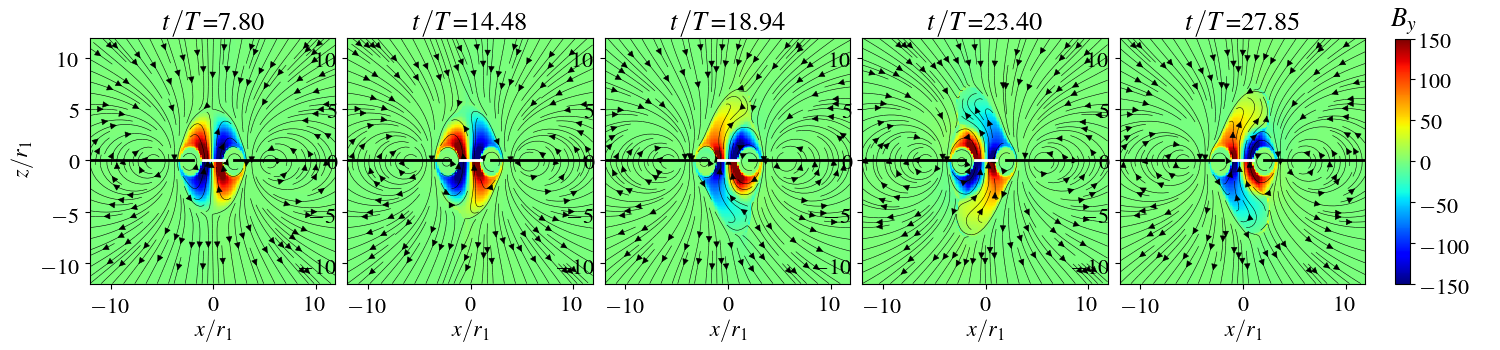}
\caption{A few snapshots from a simulation where the membrane has resistivity $\eta\equiv R/4\pi=1$, the disk current distribution has a positive current loop $j_1$ located at $r=2.0r_1$ and a negative current loop $j_2$ located at $r=5.33r_1$, with $j_2=0.7j_1$. The membrane angular velocity satisfies $\Omega r_1/c=0.84$. Stream lines show the magnetic field component on the $x-z$ plane and color shows $B_y$. }\label{fig:eta1_slice}
\end{figure*}

In what follows, we consider in more detail the case corresponding to a black hole: $\eta\equiv R/4\pi=1$. Depending on the current distribution in the disk, there are a few different regimes. 

When the inner current loop is sufficiently weak compared to the outer one (small $\alpha$ in our setup), we get a final state that is an axisymmetric, disk-hole linking, closed configuration, with zero angular velocity (since $\Omega_d=0$) as we have seen in the last column of Figure \ref{fig:alpha_0.9slice}. This is very similar to what we found in Paper I: the closed region remains sub-Alfv\'{e}nic and appears to be stable. 

On the other hand, when the inner current loop is too strong (large $\alpha$ in our setup), the resulting zero-angular-velocity configuration will develop a toroidal magnetic field that is too strong to be confined by the pressure provided by the field from the outer disk. The inner closed loop will be able to break out and open up eventually, reaching an angular velocity that is about half the angular velocity of the black hole, as required by the membrane boundary condition. 

In the intermediate $\alpha$ range, we see the development of the $m=1$ kink mode. Figure \ref{fig:eta1_slice} shows a few snapshots of the kink mode from a particular simulation where the disk current distribution has a positive loop $j_1$ located at $r=2.0r_1$ and a negative loop $j_2$ located at $r=5.33r_1$, with $j_2=0.7j_1$. The development of the kink is very similar to the case corresponding to the first row of Figure \ref{fig:eta0slice}.

\subsection{Effects of the accretion disk rotation} \label{subsec:disk_rotation}
\begin{figure*}
    \centering
    \includegraphics[width=\textwidth]{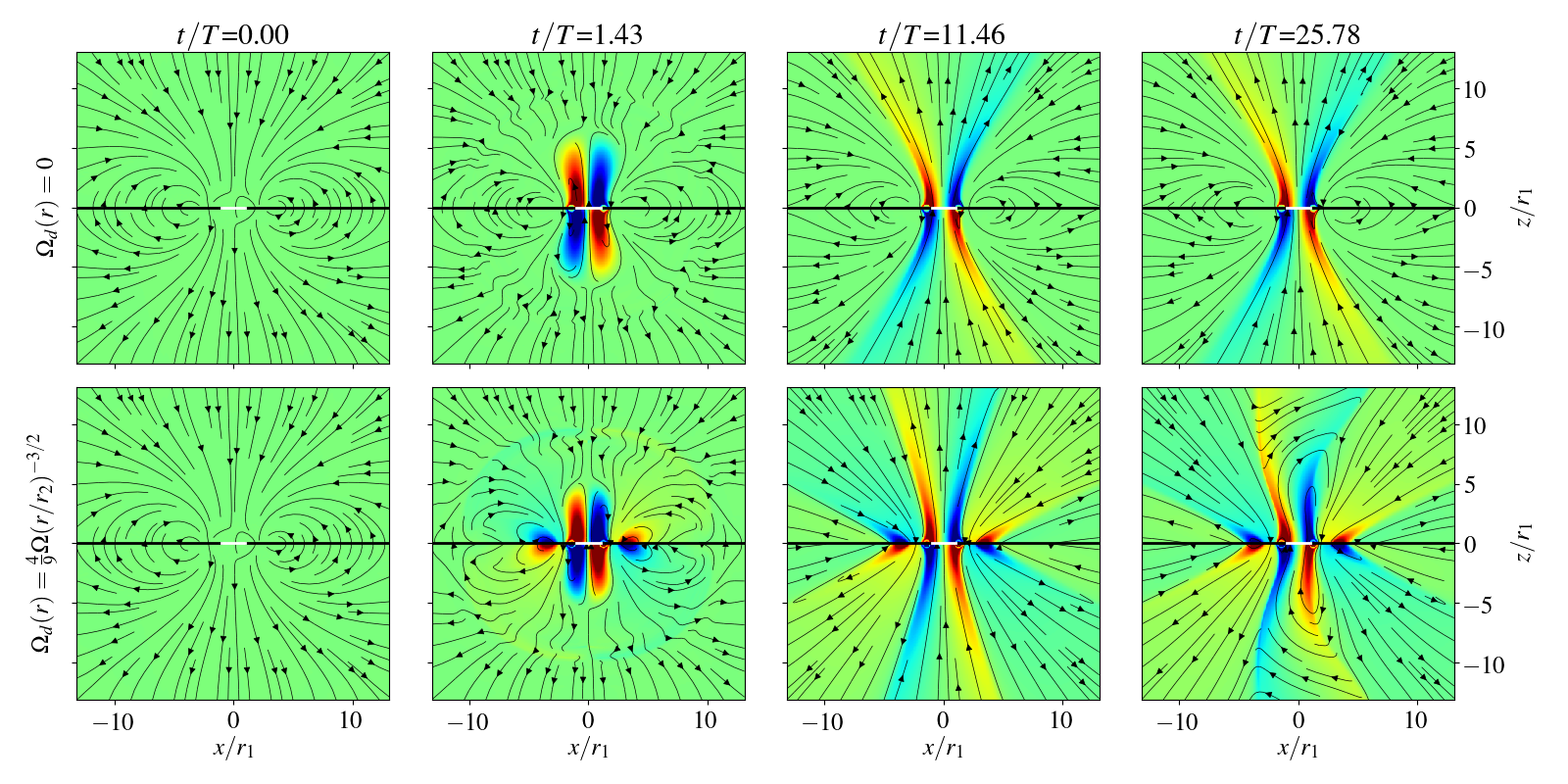}
    \caption{A comparison between the cases where the accretion disk is rotating or not rotating. In both cases, the central object is perfectly conducting, and rotating with an angular velocity such that $\Omega r_1/c=0.9$. The disk has $r_2=1.08r_1$ and the current distribution satisfies $\alpha=0.7$. In the first row, the accretion disk is not rotating: $\Omega_d(r)=0$. This is the same case as the last row in Figure \ref{fig:eta0slice}. In the bottom row, the disk is rotating with a Keplerian profile: $\Omega_d(r)=f\Omega(r/r_2)^{-3/2}$, where $f=4/9$ as an example. Snapshots at a few time points are shown. The lines correspond to the in-plane magnetic field and color indicates the out-of-plane magnetic field. In the last panel of the bottom row, we can see that the kink instability is developing.}
    \label{fig:wdslice}
\end{figure*}

So far we have set the accretion disk at rest, with $\Omega_d(r)=0$. In reality, the accretion disk is orbiting around the black hole with Keplerian angular velocity, and the shear motion could affect the evolution of the magnetic field on the accretion disk. In figure \ref{fig:wdslice} we show such an example. We take the case corresponding to the last row in Figure \ref{fig:eta0slice}, and now allow the accretion disk to rotate with angular velocity $\Omega_d(r)=f\Omega(r/r_2)^{-3/2}$, where we set $f<1$ to model a disk rotating slower than the black hole. Different $f$ values show qualitatively similar results. The rotation of the disk causes the field loop on the accretion disk to open up, launching Poynting flux/magnetized wind that flows out from every radius of the disk. If an axisymmetric constraint were imposed, at large radii the field would asymptote to a monopole configuration with alternating polarities. The interesting fact here is that the disk wind now imposes more pressure on the central ``jet'' column: the opening angle of the column is reduced and it is easier for the kink instability to set in. This again lends support to our basic conclusion that strong confinement from the disk field/disk wind makes the jet more prone to kink instability when the field coherence length scale is small.

\section{On the energy extraction and dissipation}\label{sec:dissipation}
\begin{figure*}
\includegraphics[width=0.8\textwidth]{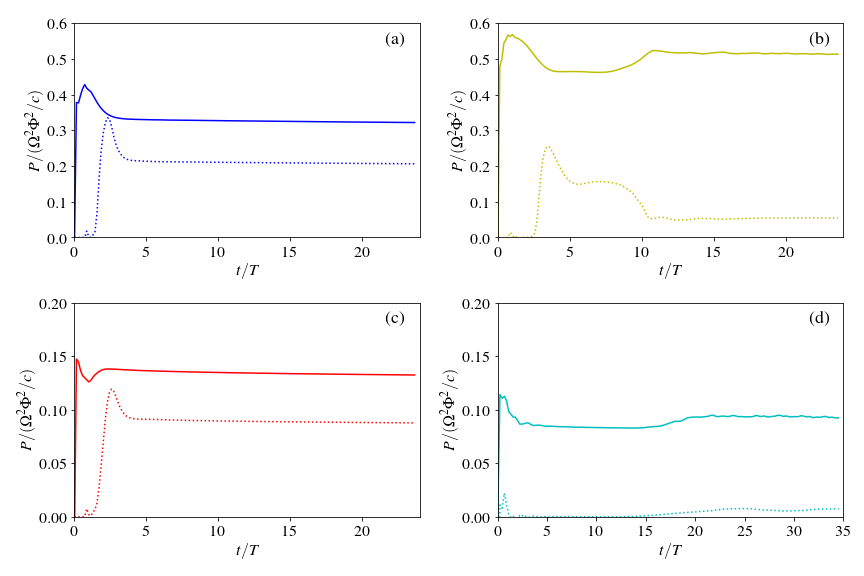}
\caption{Solid lines show the total power output from the central membrane as a function of time; dotted lines show the amount of power that goes out through a spherical surface at large radius $r_m=6r_1$. Four different cases are shown. (a): The central object is a perfect conductor and the field reaches a steady open configuration (initial current distribution in the disk has $\alpha=0.7$, $r_2=1.08r_1$, corresponding to the last row of Figure \ref{fig:eta0slice}). (b): The central object is a perfect conductor; the field first opens up, then gets destroyed by the kink instability ($\eta=0$, $\alpha=0.2$, $r_2=1.08r_1$, corresponding to the second row of Figure \ref{fig:eta0slice}). (c): the membrane has resistivity $\eta=1$, and the field reaches a steady open configuration (initial current distribution in the disk has $\alpha=1.5$, $r_2=1.08r_1$). (d): the membrane has resistivity $\eta=1$, and the field undergoes kink instability, unable to form an outflow (the initial disk current distribution has a positive current loop $j_1$ located at $r=2.0r_1$ and a negative current loop $j_2$ located at $r=5.33r_1$, with $j_2=0.7j_1$, corresponding to Figure \ref{fig:eta1_slice}). The first three cases have $\Omega r_1/c=0.9$ while the last case has $\Omega r_1/c=0.84$. It is obvious that when the field reaches a steady, open configuration, there is a significant amount of Poynting flux going out to infinity (cases a and c), but when the outflow is destroyed by kink instability, very little Poynting flux goes out at large distances (cases b and d).}\label{fig:poyntingflux}
\end{figure*}

We can directly measure the energy extraction from the central rotating membrane by integrating the Poynting flux over the surface of the membrane. Figure \ref{fig:poyntingflux} shows four examples of power output as a function of time. The total power is normalized to the fiducial value $P_0=\Omega^2\Phi^2/c$, where $\Phi$ is the magnetic flux threading the membrane at $t=0$. We can see that the power output from the membrane is indeed on the order of $P_0$. The actual prefactor  depends on the resistivity of the membrane and the way the magnetic flux is distributed on the membrane. Blue line and red line both correspond to final steady, open configurations, but when the membrane has significant resistivity (red line), the power output becomes much smaller. For the case of the yellow line ($\eta=0$), the increase around $t/T=10$ corresponds to the fully developed kink mode: although the field lines are fixed to the perfectly conducting central object and rotate at the same angular velocity, the kink mode changes the current distribution on the field lines, thus changing the Poynting flux as well. In the case of the cyan line (black hole, $\eta=1$), around $t/T\approx20$ the kink instability has fully developed; we also see a slight increase in the total power output from the membrane.\footnote{For a stable, closed configuration similar to the last column of Figure \ref{fig:alpha_0.9slice}, based on our results from Paper I, the energy extraction rate $P\propto \omega I$, where $\omega$ is the angular velocity of the field line and $I$ is the poloidal current that flows along the field lines threading the black hole, we should have $P=0$ as $\omega=0$. In reality we see small amounts of energy extraction, part of which gets dissipated volumetrically in the closed region, while the majority gets dissipated along the field lines connecting to the edge of the membrane where a boundary current layer lies.}

\begin{figure*}
\includegraphics[width=\textwidth]{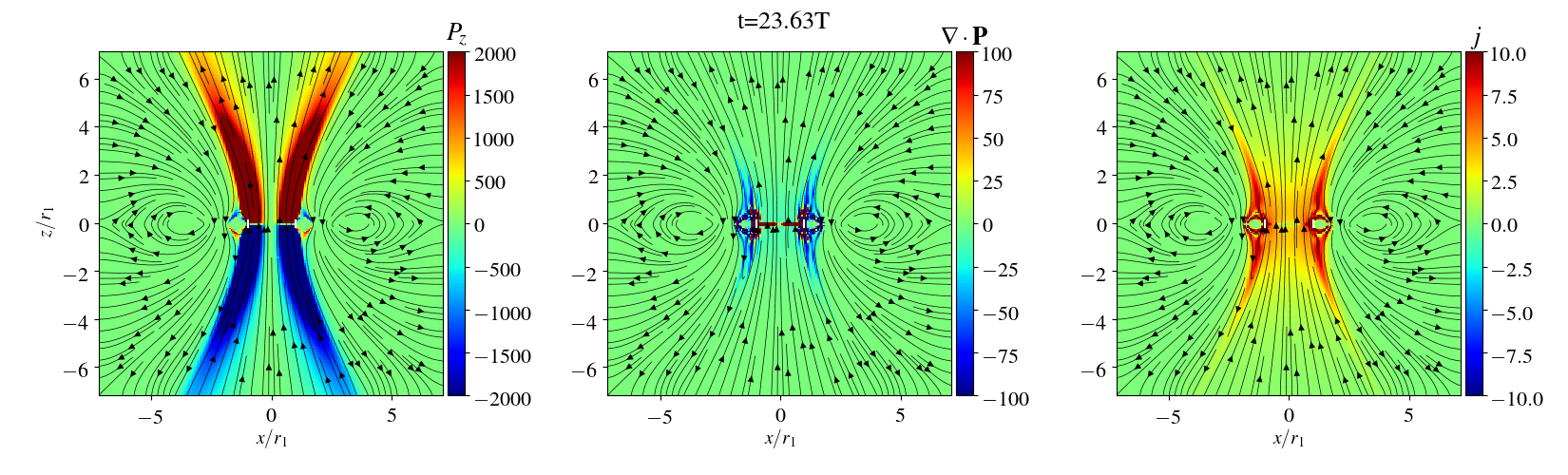}
\caption{Left panel: $z$-component of the Poynting flux (arbitrary units) for the case where the membrane has zero resistivity and the final state is an axisymmetric, open configuration, corresponding to the bottom row of Figure \ref{fig:eta0slice}. Streamlines show the poloidal magnetic field. Middle panel: the divergence of the Poynting vector, showing where the dissipation happens. Right panel: the magnitude of the current density.}\label{fig:divP_alpha0.7}
\end{figure*}

\begin{figure*}
\includegraphics[width=\textwidth]{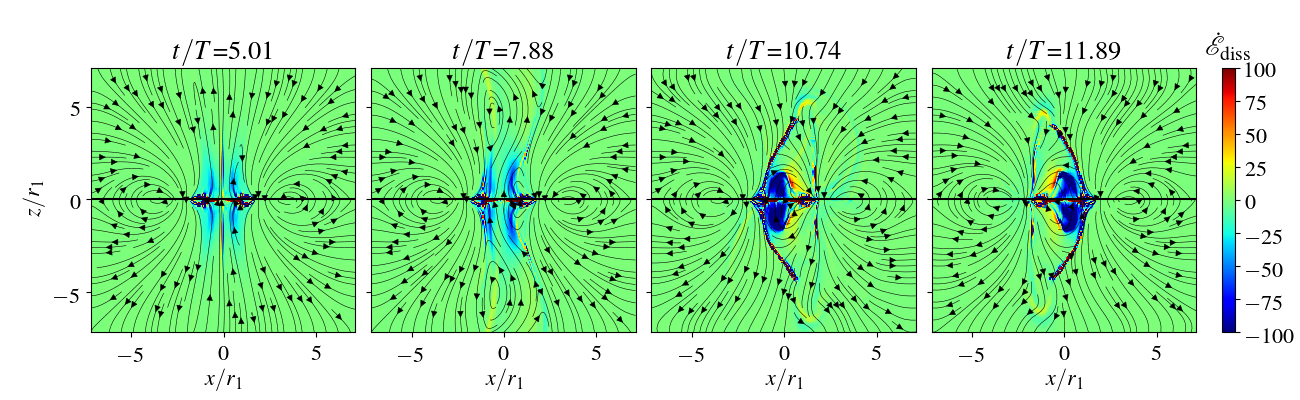}\\
\includegraphics[width=\textwidth]{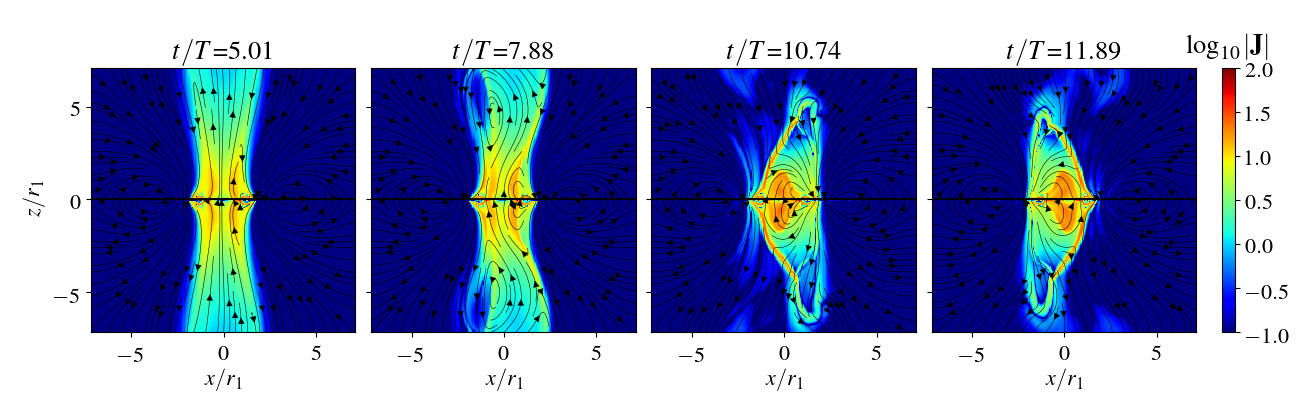}\\
\includegraphics[width=\textwidth]{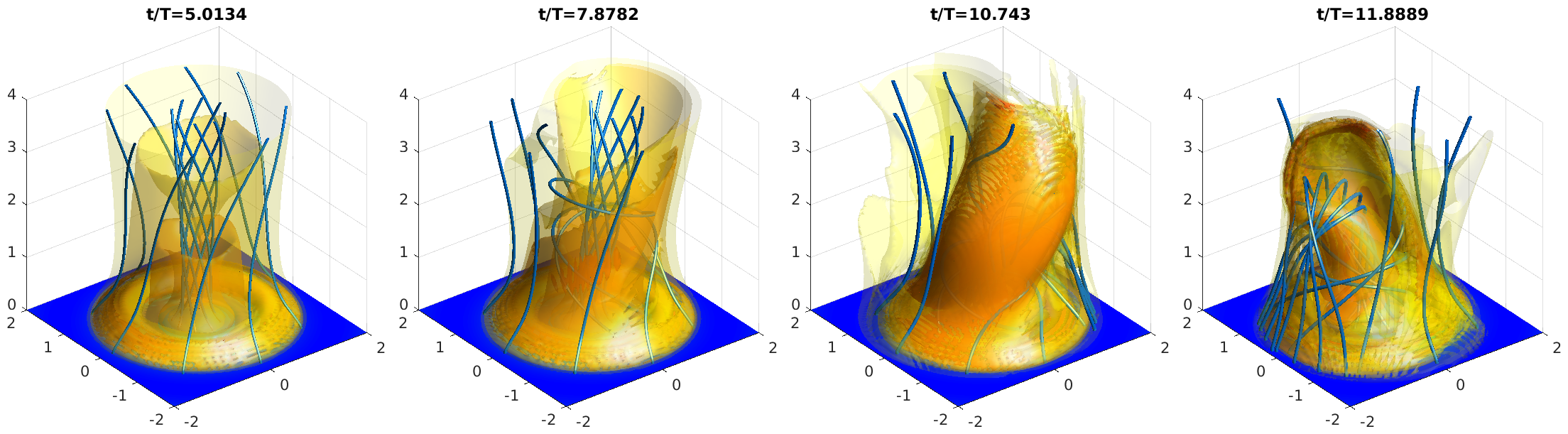}
\caption{Dissipation sites for the case where the membrane has zero resistivity and the disk current distribution satisfies $\alpha=0.2$, $r_2=1.08r_1$, corresponding to the second row of Figure \ref{fig:eta0slice}. Top row: $x-z$ slices of the numerical dissipation rate as determined by Euqation (\ref{eq:dissipation}), overlaid on the magnetic field. Second row: $x-z$ slices of current density, overlaid on the magnetic field. Bottom row: 3D rendering of the current density isosurfaces (redder color means higher current density), shown together with a bunch of magnetic field lines (similar to Figure \ref{fig:alpha0.2eta0-3d}).}\label{fig:alpha0.2eta0.0-dissipation}
\end{figure*}

The energy extracted from the central membrane may have three outcomes: (1) flowing along the open field lines toward infinity; (2) flowing along closed field lines to the disk; (3) dissipated in the magnetosphere. Since our accretion disk is perfectly conducting and non-rotating, which acts like a reflecting wall, there is no Poynting flux entering the disk, namely, component (2) should be zero. Component (1) becomes negligible when most of the field lines are closed, or open flux bundle gets destroyed by the kink mode (dotted lines in Figure \ref{fig:poyntingflux}). As a result, in these cases most of the extracted rotational energy gets dissipated in the magnetosphere. In what follows we would like to get a sense of where the dissipation tends to take place in the magnetosphere.
When force-free condition is satisfied, dissipation does not happen because $\pmb{E}\cdot\pmb{J}=0$ is ensured by the force-free condition. However, as we have seen, the system may evolve toward a state where the force-free condition gets violated in certain regions, and dissipation has to happen there. In our current simulation scheme, the dissipation is numerical in origin. Force-free gets violated when $\pmb{E}\cdot\pmb{B}\neq0$ or $E>B$: we then directly reduce the corresponding component of $\pmb{E}$ to restore it back to force-free. This can also be regarded as efficient radiation that removes the electromagnetic energy from the system.  

Based on the energy conservation law, we have
\begin{equation}\label{eq:dissipation}
\frac{\partial}{\partial t}\frac{B^2+E^2}{8\pi}+\frac{c}{4\pi}\nabla\cdot(\pmb{E}\times\pmb{B})=\dot{\mathcal{E}}_{\rm diss}
\end{equation}
Writing the energy density of the electromagnetic field as $\mathcal{E}=(B^2+E^2)/(8\pi)$,
and the Poynting vector as $\pmb{P}=\pmb{E}\times\pmb{B}c/(4\pi)$,
we get
\begin{equation}\label{eq:dissipation1}
\partial \mathcal{E}/\partial t+\nabla\cdot\pmb{P}=\dot{\mathcal{E}}_{\rm diss}.
\end{equation}
For ideal force-free, $\dot{\mathcal{E}}_{\rm diss}=0$, but it will be nonzero in regions where force-free condition is violated. We can directly calculate $\dot{\mathcal{E}}_{\rm diss}$ using the left hand side of Equation (\ref{eq:dissipation1}) to see where the dissipation happens.

Figure \ref{fig:divP_alpha0.7} shows an example for the case where the membrane is perfectly conducting and the system eventually reaches a steady, axisymmetric open configuration, corresponding to the last row of Figure \ref{fig:eta0slice}. In this case, $\dot{\mathcal{E}}_{\rm diss}=\nabla\cdot\pmb{P}$. By comparing the plot of $\dot{\mathcal{E}}_{\rm diss}$ with the plot of current density, we see that the current sheet at the jet wall is the main site for dissipation. 

Figure \ref{fig:alpha0.2eta0.0-dissipation} shows another example where the evolution is highly dynamic and asymmetric. This case corresponds to the second row of Figure \ref{fig:eta0slice} where the kink instability fully destroys the outflow. We can see again that the dissipation sites coincide with places where the current is strong. In the first column, a transient state with field lines attached to the central star fully open, there is a current layer separating the field lines from the star and those from the disk, which have different polarities. The current distribution and dissipation are similar to those in Figure \ref{fig:divP_alpha0.7}. When the kink has fully developed, we can see a strong current layer wrapping around the closed flux bundle. This is where the separatrix lies, and where the field lines reconnect and alternate between closed/open configurations. Indeed these appear to be strong dissipation sites. 
However, in order to accurately account for the energy dissipation rate, we need to include more physical dissipation mechanisms. This will be studied in a future work. 


\section{Discussion and Conclusions}\label{sec:dicussion_conclusion}

In this paper, we use a simple setup in special relativistic force-free framework to study the dynamics of small scale flux tubes connecting the central black hole/star and the surrounding thin accretion disk. For a perfectly conducting star, the field that anchors on the star will always extract significant amount of energy from the star, but whether an outflow can be formed depends on the competition between the pressure of the twisted field in the magnetic tower and the external field that confines it. Strong confinement from the surrounding field could suppress the formation of an outflow and lead to the dissipation of most of the extracted energy near the central star. Weak confinement, on the other hand, facilitates the formation of a relatively stable outflow of Poynting flux with a wide opening angle as the magnetic twist builds up. 
For a black hole, since it behaves like a resistive membrane and the field lines can slip on it, there is an additional regime where small scale flux tubes can be well confined and little energy is extracted or dissipated. In the intermediate confinement regime, the black hole twists up the field and forms a magnetic tower, which gets disrupted by $m=1$ (kink) instability. In this particularly interesting regime, energy extraction increases but most of it is dissipated in the closed region disturbed by the instability. 
This demonstrates that, in principle, with the field configuration near the black hole being filled by small scale flux tubes, it is possible that the rotational energy is extracted from the black hole but gets dissipated relatively nearby, instead of forming an outflow in the form of a Blandford-Znajek jet. This may be relevant for the heating of the compact X-ray emitting corona in Seyfert Galaxies.

We have seen that in the regime where the kink instability disrupts the outflow, a significant fraction of the extracted rotational energy $P\sim P_0=\Omega^2\Phi^2/c$ gets dissipated in the magnetosphere, especially at the coherent current sheets formed by the kink instability. These current sheets also rotate around the central compact object, with a rate that is determined by the kink time scale and typically some fraction of the angular velocity of the black hole. This could lead to quasi-periodic signatures in the observed radiation, especially if the radiation mechanism at the current sheet is to some extent beamed. However, in an actual accretion disk the flux tubes may be much more irregular and variable, leading to more stochasticity in the observed signal. 

The main dissipation mechanism is likely to be reconnection at these current sheets. Since the environment around the black hole is filled by soft optical/UV photons from the accretion disk, particles accelerated in the reconnection process will be quickly cooled through inverse Compton scatterings and may also initiate pair cascades on the soft photons. The reconnection outflow may be strongly Compton dragged, and the reconnection proceeds in a regime as described by \citet{2017ApJ...850..141B} where the spectrum is dominated by bulk Comptonization from the reconnection outflow and has a high energy cutoff near the pair production threshold.

To get an estimation of the dissipation rate in realistic systems, let us consider a typical Seyfert Galaxy, MCG-6-30-15, which has a supermassive black hole of mass $M=2\times 10^6M_{\odot}$, bolometric luminosity $L\sim 8\times 10^{43}\,\rm{erg}\,\rm{s}^{-1}\sim 0.4L_{\rm Edd}$, and X-ray luminosity $L_X\sim2\times10^{43}\,\rm{erg}\,\rm{s}^{-1}\sim 0.1L_{\rm Edd}$\citep{1997MNRAS.291..403R}. Measuring $B$ in terms of the characteristic field $B_{\rm Edd}\equiv\sqrt{8\pi p_{\rm Edd}}=\sqrt{2L_{\rm Edd}/3cr_g^2}=\sqrt{m_pc^2/r_e^2r_g}=3.6\times10^5M_6^{-1/2}$ G \citep{phinney_theory_1983}, we get 
\begin{equation}
P_0\sim a^2\left(\frac{B}{B_{\rm Edd}}\right)^2L_{\rm Edd}.
\end{equation}
If the magnetic field near the black hole reaches a value $B\sim B_{\rm Edd}$, then the amount of power dissipated due to the kink instability will be enough to account for all the X-ray emission. 

The kind of relatively strong magnetic field with small coherence length scales may be a generic outcome of MRI in the accretion disk, or associated with hot spots on the accretion disk as the flux tubes emerge buoyantly similar to solar magnetosphere. As we have seen from GRMHD simulations, when the accretion disk is threaded by a strong net magnetic flux, it tends to become magnetically arrested and form a powerful jet \citep{McKinney2009MNRAS.394L.126M,Avara2016MNRAS.462..636A}; however, in situations where the net magnetic flux is small, the disk magnetic field may be dominated by the locally amplified components. MRI and the subsequent buoyancy or gravitational instabilities can produce fields with alternating polarities \citep[e.g.,][]{2010ApJ...713...52D,2018ApJ...857...34Z}; they may also be quite asymmetric. As these structures are fed to the black hole, we may see continuous flux tube inflation and kink instability happening on a much more stochastic basis. We have seen from the simulations that the kink instability tends to happen when the confinement from the surrounding field is strong; this confinement can be produced by magnetic field from currents in the outer disk, or a disk wind that exerts pressure towards the central axis region.

We have so far approximated the black hole as a resistive membrane in a flat spacetime model. Whether the other GR effects are important will be tested using a time-dependent, GR force-free code in the future. Meanwhile, we used an ad hoc way to deal with the current sheet where force-free condition is violated. When the non-force-free regions have negligible volume, it can be shown that the overall dynamics does not depend sensitively on the dissipation mechanisms in these boundary layers, and force-free evolution gives a reasonable zero-th order picture \citep{2015PhRvL.115i5002E,2016ApJ...817...89Z,2016ApJ...826..115N} \footnote{Force-free evolution becomes problematic if a large volume violates the force-free condition \citep{2017JPlPh..83f6301L,2018arXiv181010493L}.}. However, to get a better estimation of the energy dissipation rate, we need to use more realistic prescriptions for the dissipation in the force-free code. Another limitation of our approach is that we use the force-free electrodynamics, neglecting the inertia of the plasma attached to the coronal field lines. In regions where the mass loading is small such that the Alfv\'{e}n speed is close to c and much larger than the sound speed, force-free is a good approximation. We can get a rough estimation of the plasma magnetization for actual astrophysical systems like MCG-6-30-15 \citep[see also similar discussion in][]{2019MNRAS.484.4920Y}. The magnetic field near the black hole may be on the order of $B\sim 0.3B_{\rm Edd}\sim 10^5$ G; the density of the electrons $n_e$ can be estimated from the Compton y parameter $y=4kT \rm{Max}(\tau_{\rm es},\tau_{\rm es}^2)/(mc^2)$>1, where $\tau_{\rm es}=n_e\sigma_T r$ is the Thomson scattering optical depth, $T$ is the temperature of the electrons, and $r$ is the size of the region. Since $\tau_{\rm es}\gtrsim1$, we get 
\begin{equation}
    n_e=6\times10^{11}y^{1/2}\left(\frac{T}{100\,\rm{keV}}\right)^{-1/2}\left(\frac{r}{10r_g}\right)^{-1}\,\rm{cm}^{-3}.
\end{equation}
So the electron magnetization is
\begin{equation}
    \sigma_e\equiv\frac{B^2}{4\pi n_e mc^2}\approx 2\times 10^3\left(\frac{T}{100\,\rm{keV}}\right)^{1/2}\left(\frac{r}{10r_g}\right)\left(\frac{B}{10^5\,\rm{G}}\right)^2.
\end{equation}
If proton number density is the same as the electron number density, the total magnetization would be order 1. But the corona emitting region may well be electron/positron pair dominated \citep{1996MNRAS.283..193Z,2017ApJ...850..141B}, meaning that the total magnetization remains high. In such situations, the force-free approximation gives a reasonable description.
However, another complication is that the mass loading could be quite large for flux tubes that are directed inward at small enough angles to the plane of the disk for which gravitational attraction will overcome centrifugal force and allow disk gas to be unstable to flow along the magnetic field lines \citep{Blandford1982MNRAS.199..883B,2019MNRAS.484.4920Y}. This could change the characteristic wave speeds along the magnetic field, affecting the criterion as to which parts of the flux tubes are more likely to break up and reconnect when the flow becomes supersonic, a process that could enhance the amount of dissipation on the axis near the black hole. A MHD approach is needed to capture such dynamics.
Finally, here we do not include the accretion disk motion and field evolution self-consistently; a better approach would be to model the disk using full MHD and allow all scales of magnetic structures to form and evolve self-consistently. These will be investigated in future works.

\section*{Acknowledgements}

We thank Omer Bromberg, Alexander Yuran Chen, Dmitri Uzdensky for helpful discussions, and the anonymous referee for useful suggestions. YY acknowledges support from the Lyman Spitzer, Jr. Postdoctoral Fellowship awarded by the Department of Astrophysical Sciences at Princeton University. AS is supported by NASA ATP grant 80NSSC18K1099 and by Simons Foundation (grant 267233). DRW is supported by NASA through Einstein Postdoctoral Fellowship grant number PF6-170160, awarded by the \textit{Chandra} X-ray Center, operated by the Smithsonian Astrophysical Observatory for NASA under contract NAS8-03060. The simulations presented in this work used computational resources supported by the PICSciE-OIT High Performance Computing Center at Princeton University.




\bibliographystyle{mnras}
\bibliography{ref}

\begin{thebibliography}{}
\makeatletter
\relax
\def\mn@urlcharsother{\let\do\@makeother \do\$\do\&\do\#\do\^\do\_\do\%\do\~}
\def\mn@doi{\begingroup\mn@urlcharsother \@ifnextchar [ {\mn@doi@}
  {\mn@doi@[]}}
\def\mn@doi@[#1]#2{\def\@tempa{#1}\ifx\@tempa\@empty \href
  {http://dx.doi.org/#2} {doi:#2}\else \href {http://dx.doi.org/#2} {#1}\fi
  \endgroup}
\def\mn@eprint#1#2{\mn@eprint@#1:#2::\@nil}
\def\mn@eprint@arXiv#1{\href {http://arxiv.org/abs/#1} {{\tt arXiv:#1}}}
\def\mn@eprint@dblp#1{\href {http://dblp.uni-trier.de/rec/bibtex/#1.xml}
  {dblp:#1}}
\def\mn@eprint@#1:#2:#3:#4\@nil{\def\@tempa {#1}\def\@tempb {#2}\def\@tempc
  {#3}\ifx \@tempc \@empty \let \@tempc \@tempb \let \@tempb \@tempa \fi \ifx
  \@tempb \@empty \def\@tempb {arXiv}\fi \@ifundefined
  {mn@eprint@\@tempb}{\@tempb:\@tempc}{\expandafter \expandafter \csname
  mn@eprint@\@tempb\endcsname \expandafter{\@tempc}}}

\bibitem[\protect\citeauthoryear{{Achterberg}}{{Achterberg}}{1996a}]{1996A&A...313.1008A}
{Achterberg} A.,  1996a, \aap, \href
  {http://adsabs.harvard.edu/abs/1996A%26A...313.1008A} {313, 1008}

\bibitem[\protect\citeauthoryear{{Achterberg}}{{Achterberg}}{1996b}]{1996A&A...313.1016A}
{Achterberg} A.,  1996b, \aap, \href
  {http://adsabs.harvard.edu/abs/1996A%26A...313.1016A} {313, 1016}

\bibitem[\protect\citeauthoryear{{Appl}, {Lery}  \& {Baty}}{{Appl}
  et~al.}{2000}]{2000A&A...355..818A}
{Appl} S.,  {Lery} T.,   {Baty} H.,  2000, \aap, \href
  {http://adsabs.harvard.edu/abs/2000A%26A...355..818A} {355, 818}

\bibitem[\protect\citeauthoryear{{Avara}, {McKinney}  \& {Reynolds}}{{Avara}
  et~al.}{2016}]{Avara2016MNRAS.462..636A}
{Avara} M.~J.,  {McKinney} J.~C.,   {Reynolds} C.~S.,  2016, \mn@doi [\mnras]
  {10.1093/mnras/stw1643}, \href
  {http://adsabs.harvard.edu/abs/2016MNRAS.462..636A} {462, 636}

\bibitem[\protect\citeauthoryear{{Beloborodov}}{{Beloborodov}}{2017}]{2017ApJ...850..141B}
{Beloborodov} A.~M.,  2017, \mn@doi [\apj] {10.3847/1538-4357/aa8f4f}, \href
  {https://ui.adsabs.harvard.edu/\#abs/2017ApJ...850..141B} {850, 141}

\bibitem[\protect\citeauthoryear{{Blandford}}{{Blandford}}{2002}]{2002luml.conf..381B}
{Blandford} R.~D.,  2002, in {Gilfanov} M.,  {Sunyeav} R.,   {Churazov} E.,
  eds, Lighthouses of the Universe: The Most Luminous Celestial Objects and
  Their Use for Cosmology. p.~381 (\mn@eprint {} {astro-ph/0202265}),
  \mn@doi{10.1007/10856495_59}

\bibitem[\protect\citeauthoryear{{Blandford} \& {Payne}}{{Blandford} \&
  {Payne}}{1982}]{Blandford1982MNRAS.199..883B}
{Blandford} R.~D.,  {Payne} D.~G.,  1982, \mn@doi [\mnras]
  {10.1093/mnras/199.4.883}, \href
  {http://adsabs.harvard.edu/abs/1982MNRAS.199..883B} {199, 883}

\bibitem[\protect\citeauthoryear{{Blandford} \& {Znajek}}{{Blandford} \&
  {Znajek}}{1977}]{Blandford1977MNRAS.179..433B}
{Blandford} R.~D.,  {Znajek} R.~L.,  1977, \mn@doi [\mnras]
  {10.1093/mnras/179.3.433}, \href
  {http://adsabs.harvard.edu/abs/1977MNRAS.179..433B} {179, 433}

\bibitem[\protect\citeauthoryear{{Bromberg} \& {Tchekhovskoy}}{{Bromberg} \&
  {Tchekhovskoy}}{2016}]{2016MNRAS.456.1739B}
{Bromberg} O.,  {Tchekhovskoy} A.,  2016, \mn@doi [\mnras]
  {10.1093/mnras/stv2591}, \href
  {http://adsabs.harvard.edu/abs/2016MNRAS.456.1739B} {456, 1739}

\bibitem[\protect\citeauthoryear{{Cerutti}, {Philippov}, {Parfrey}  \&
  {Spitkovsky}}{{Cerutti} et~al.}{2015}]{2015MNRAS.448..606C}
{Cerutti} B.,  {Philippov} A.,  {Parfrey} K.,   {Spitkovsky} A.,  2015, \mn@doi
  [\mnras] {10.1093/mnras/stv042}, \href
  {http://adsabs.harvard.edu/abs/2015MNRAS.448..606C} {448, 606}

\bibitem[\protect\citeauthoryear{{Chartas}, {Kochanek}, {Dai}, {Poindexter}  \&
  {Garmire}}{{Chartas} et~al.}{2009}]{Chartas2009ApJ...693..174C}
{Chartas} G.,  {Kochanek} C.~S.,  {Dai} X.,  {Poindexter} S.,   {Garmire} G.,
  2009, \mn@doi [\apj] {10.1088/0004-637X/693/1/174}, \href
  {http://adsabs.harvard.edu/abs/2009ApJ...693..174C} {693, 174}

\bibitem[\protect\citeauthoryear{{Davis}, {Stone}  \& {Pessah}}{{Davis}
  et~al.}{2010}]{2010ApJ...713...52D}
{Davis} S.~W.,  {Stone} J.~M.,   {Pessah} M.~E.,  2010, \mn@doi [\apj]
  {10.1088/0004-637X/713/1/52}, \href
  {http://adsabs.harvard.edu/abs/2010ApJ...713...52D} {713, 52}

\bibitem[\protect\citeauthoryear{{East}, {Zrake}, {Yuan}  \&
  {Blandford}}{{East} et~al.}{2015}]{2015PhRvL.115i5002E}
{East} W.~E.,  {Zrake} J.,  {Yuan} Y.,   {Blandford} R.~D.,  2015, \mn@doi
  [Physical Review Letters] {10.1103/PhysRevLett.115.095002}, \href
  {http://adsabs.harvard.edu/abs/2015PhRvL.115i5002E} {115, 095002}

\bibitem[\protect\citeauthoryear{{Gruzinov}}{{Gruzinov}}{1999}]{1999astro.ph..2288G}
{Gruzinov} A.,  1999, preprint, \href
  {https://ui.adsabs.harvard.edu/#abs/1999astro.ph..2288G} {pp
  astro--ph/9902288} (\mn@eprint {arXiv} {astro-ph/9902288})

\bibitem[\protect\citeauthoryear{{Kara}, {Alston}, {Fabian}, {Cackett},
  {Uttley}, {Reynolds}  \& {Zoghbi}}{{Kara}
  et~al.}{2016}]{Kara2016MNRAS.462..511K}
{Kara} E.,  {Alston} W.~N.,  {Fabian} A.~C.,  {Cackett} E.~M.,  {Uttley} P.,
  {Reynolds} C.~S.,   {Zoghbi} A.,  2016, \mn@doi [\mnras]
  {10.1093/mnras/stw1695}, \href
  {http://adsabs.harvard.edu/abs/2016MNRAS.462..511K} {462, 511}

\bibitem[\protect\citeauthoryear{{Li}, {Zrake}  \& {Beloborodov}}{{Li}
  et~al.}{2018}]{2018arXiv181010493L}
{Li} X.,  {Zrake} J.,   {Beloborodov} A.~M.,  2018, arXiv e-prints, \href
  {https://ui.adsabs.harvard.edu/abs/2018arXiv181010493L} {p. arXiv:1810.10493}

\bibitem[\protect\citeauthoryear{{Lynden-Bell}}{{Lynden-Bell}}{1996}]{1996MNRAS.279..389L}
{Lynden-Bell} D.,  1996, \mn@doi [\mnras] {10.1093/mnras/279.2.389}, \href
  {http://adsabs.harvard.edu/abs/1996MNRAS.279..389L} {279, 389}

\bibitem[\protect\citeauthoryear{{Lyubarskij}}{{Lyubarskij}}{1992}]{1992SvAL...18..356L}
{Lyubarskij} Y.~E.,  1992, Soviet Astronomy Letters, \href
  {http://adsabs.harvard.edu/abs/1992SvAL...18..356L} {18, 356}

\bibitem[\protect\citeauthoryear{{Lyutikov}, {Sironi}, {Komissarov}  \&
  {Porth}}{{Lyutikov} et~al.}{2017}]{2017JPlPh..83f6301L}
{Lyutikov} M.,  {Sironi} L.,  {Komissarov} S.~S.,   {Porth} O.,  2017, \mn@doi
  [Journal of Plasma Physics] {10.1017/S0022377817000629}, \href
  {http://adsabs.harvard.edu/abs/2017JPlPh..83f6301L} {83, 635830601}

\bibitem[\protect\citeauthoryear{{McKinney} \& {Blandford}}{{McKinney} \&
  {Blandford}}{2009}]{McKinney2009MNRAS.394L.126M}
{McKinney} J.~C.,  {Blandford} R.~D.,  2009, \mn@doi [\mnras]
  {10.1111/j.1745-3933.2009.00625.x}, \href
  {http://adsabs.harvard.edu/abs/2009MNRAS.394L.126M} {394, L126}

\bibitem[\protect\citeauthoryear{{Michel}}{{Michel}}{1973}]{1973ApJ...180L.133M}
{Michel} F.~C.,  1973, \mn@doi [\apjl] {10.1086/181169}, \href
  {http://adsabs.harvard.edu/abs/1973ApJ...180L.133M} {180, L133}

\bibitem[\protect\citeauthoryear{{Morgan}, {Kochanek}, {Dai}, {Morgan}  \&
  {Falco}}{{Morgan} et~al.}{2008}]{Morgan2008ApJ...689..755M}
{Morgan} C.~W.,  {Kochanek} C.~S.,  {Dai} X.,  {Morgan} N.~D.,   {Falco} E.~E.,
   2008, \mn@doi [\apj] {10.1086/592767}, \href
  {http://adsabs.harvard.edu/abs/2008ApJ...689..755M} {689, 755}

\bibitem[\protect\citeauthoryear{{Mosquera}, {Kochanek}, {Chen}, {Dai},
  {Blackburne}  \& {Chartas}}{{Mosquera}
  et~al.}{2013}]{Mosquera2013ApJ...769...53M}
{Mosquera} A.~M.,  {Kochanek} C.~S.,  {Chen} B.,  {Dai} X.,  {Blackburne}
  J.~A.,   {Chartas} G.,  2013, \mn@doi [\apj] {10.1088/0004-637X/769/1/53},
  \href {http://adsabs.harvard.edu/abs/2013ApJ...769...53M} {769, 53}

\bibitem[\protect\citeauthoryear{{Nalewajko}, {Zrake}, {Yuan}, {East}  \&
  {Blandford}}{{Nalewajko} et~al.}{2016}]{2016ApJ...826..115N}
{Nalewajko} K.,  {Zrake} J.,  {Yuan} Y.,  {East} W.~E.,   {Blandford} R.~D.,
  2016, \mn@doi [\apj] {10.3847/0004-637X/826/2/115}, \href
  {http://adsabs.harvard.edu/abs/2016ApJ...826..115N} {826, 115}

\bibitem[\protect\citeauthoryear{{Parfrey}, {Giannios}  \&
  {Beloborodov}}{{Parfrey} et~al.}{2015}]{Parfrey2015MNRAS.446L..61P}
{Parfrey} K.,  {Giannios} D.,   {Beloborodov} A.~M.,  2015, \mn@doi [\mnras]
  {10.1093/mnrasl/slu162}, \href
  {http://adsabs.harvard.edu/abs/2015MNRAS.446L..61P} {446, L61}

\bibitem[\protect\citeauthoryear{Phinney}{Phinney}{1983}]{phinney_theory_1983}
Phinney E.~S.,  1983, PhD thesis, University of Cambridge

\bibitem[\protect\citeauthoryear{{Porth} \& {Komissarov}}{{Porth} \&
  {Komissarov}}{2015}]{2015MNRAS.452.1089P}
{Porth} O.,  {Komissarov} S.~S.,  2015, \mn@doi [\mnras]
  {10.1093/mnras/stv1295}, \href
  {http://adsabs.harvard.edu/abs/2015MNRAS.452.1089P} {452, 1089}

\bibitem[\protect\citeauthoryear{{Reis} \& {Miller}}{{Reis} \&
  {Miller}}{2013}]{ReisMiller2013ApJ...769L...7R}
{Reis} R.~C.,  {Miller} J.~M.,  2013, \mn@doi [\apjl]
  {10.1088/2041-8205/769/1/L7}, \href
  {http://adsabs.harvard.edu/abs/2013ApJ...769L...7R} {769, L7}

\bibitem[\protect\citeauthoryear{{Reynolds}, {Ward}, {Fabian}  \&
  {Celotti}}{{Reynolds} et~al.}{1997}]{1997MNRAS.291..403R}
{Reynolds} C.~S.,  {Ward} M.~J.,  {Fabian} A.~C.,   {Celotti} A.,  1997,
  \mn@doi [\mnras] {10.1093/mnras/291.3.403}, \href
  {http://adsabs.harvard.edu/abs/1997MNRAS.291..403R} {291, 403}

\bibitem[\protect\citeauthoryear{{Spitkovsky}}{{Spitkovsky}}{2006}]{2006ApJ...648L..51S}
{Spitkovsky} A.,  2006, \mn@doi [\apjl] {10.1086/507518}, \href
  {http://adsabs.harvard.edu/abs/2006ApJ...648L..51S} {648, L51}

\bibitem[\protect\citeauthoryear{Thorne, Price  \& Macdonald}{Thorne
  et~al.}{1986}]{thorne_black_1986}
Thorne K.~S.,  Price R.~H.,   Macdonald D.~A.,  eds, 1986, Black holes : the
  membrane paradigm.
New Haven: Yale University Press

\bibitem[\protect\citeauthoryear{{Uttley}, {Cackett}, {Fabian}, {Kara}  \&
  {Wilkins}}{{Uttley} et~al.}{2014}]{Uttley2014A&ARv..22...72U}
{Uttley} P.,  {Cackett} E.~M.,  {Fabian} A.~C.,  {Kara} E.,   {Wilkins} D.~R.,
  2014, \mn@doi [\aapr] {10.1007/s00159-014-0072-0}, \href
  {http://adsabs.harvard.edu/abs/2014A%26ARv..22...72U} {22, 72}

\bibitem[\protect\citeauthoryear{{Uzdensky}}{{Uzdensky}}{2005}]{Uzdensky2005ApJ...620..889U}
{Uzdensky} D.~A.,  2005, \mn@doi [\apj] {10.1086/427180}, \href
  {http://adsabs.harvard.edu/abs/2005ApJ...620..889U} {620, 889}

\bibitem[\protect\citeauthoryear{{Uzdensky} \& {Goodman}}{{Uzdensky} \&
  {Goodman}}{2008}]{UzdenskyGoodman2008ApJ...682..608U}
{Uzdensky} D.~A.,  {Goodman} J.,  2008, \mn@doi [\apj] {10.1086/588812}, \href
  {http://adsabs.harvard.edu/abs/2008ApJ...682..608U} {682, 608}

\bibitem[\protect\citeauthoryear{{Uzdensky}, {K{\"o}nigl}  \&
  {Litwin}}{{Uzdensky} et~al.}{2002}]{2002ApJ...565.1191U}
{Uzdensky} D.~A.,  {K{\"o}nigl} A.,   {Litwin} C.,  2002, \mn@doi [\apj]
  {10.1086/324720}, \href {http://adsabs.harvard.edu/abs/2002ApJ...565.1191U}
  {565, 1191}

\bibitem[\protect\citeauthoryear{{Wilkins} \& {Fabian}}{{Wilkins} \&
  {Fabian}}{2011}]{Wilkins2011MNRAS.414.1269W}
{Wilkins} D.~R.,  {Fabian} A.~C.,  2011, \mn@doi [\mnras]
  {10.1111/j.1365-2966.2011.18458.x}, \href
  {http://adsabs.harvard.edu/abs/2011MNRAS.414.1269W} {414, 1269}

\bibitem[\protect\citeauthoryear{{Wilkins} \& {Gallo}}{{Wilkins} \&
  {Gallo}}{2015}]{Wilkins2015MNRAS.449..129W}
{Wilkins} D.~R.,  {Gallo} L.~C.,  2015, \mn@doi [\mnras]
  {10.1093/mnras/stv162}, \href
  {http://adsabs.harvard.edu/abs/2015MNRAS.449..129W} {449, 129}

\bibitem[\protect\citeauthoryear{{Yuan} \& {Narayan}}{{Yuan} \&
  {Narayan}}{2014}]{2014ARA&A..52..529Y}
{Yuan} F.,  {Narayan} R.,  2014, \mn@doi [\araa]
  {10.1146/annurev-astro-082812-141003}, \href
  {http://adsabs.harvard.edu/abs/2014ARA%26A..52..529Y} {52, 529}

\bibitem[\protect\citeauthoryear{{Yuan}, {Blandford}  \& {Wilkins}}{{Yuan}
  et~al.}{2019}]{2019MNRAS.484.4920Y}
{Yuan} Y.,  {Blandford} R.~D.,   {Wilkins} D.~R.,  2019, \mn@doi [\mnras]
  {10.1093/mnras/stz332}, \href
  {http://adsabs.harvard.edu/abs/2019MNRAS.484.4920Y} {484, 4920}

\bibitem[\protect\citeauthoryear{{Zdziarski}, {Johnson}  \&
  {Magdziarz}}{{Zdziarski} et~al.}{1996}]{1996MNRAS.283..193Z}
{Zdziarski} A.~A.,  {Johnson} W.~N.,   {Magdziarz} P.,  1996, \mn@doi [\mnras]
  {10.1093/mnras/283.1.193}, \href
  {http://adsabs.harvard.edu/abs/1996MNRAS.283..193Z} {283, 193}

\bibitem[\protect\citeauthoryear{{Zhu} \& {Stone}}{{Zhu} \&
  {Stone}}{2018}]{2018ApJ...857...34Z}
{Zhu} Z.,  {Stone} J.~M.,  2018, \mn@doi [\apj] {10.3847/1538-4357/aaafc9},
  \href {https://ui.adsabs.harvard.edu/\#abs/2018ApJ...857...34Z} {857, 34}

\bibitem[\protect\citeauthoryear{{Zrake} \& {East}}{{Zrake} \&
  {East}}{2016}]{2016ApJ...817...89Z}
{Zrake} J.,  {East} W.~E.,  2016, \mn@doi [\apj] {10.3847/0004-637X/817/2/89},
  \href {https://ui.adsabs.harvard.edu/abs/2016ApJ...817...89Z} {817, 89}

\makeatother
\end{thebibliography}




\appendix

\section{Steady state tests of the membrane boundary condition}\label{sec:membrane_test}
To test the membrane boundary condition, we first apply it to a relaxation scheme that solves the steady-state, Grad-Shafranov equation in flat spacetime (we have taken $c=1$):
\begin{equation}\label{eq:GS}
\begin{split}
\left(-1+\omega ^2 r^2\sin ^2\theta \right)\nabla ^2\Psi +\frac{2}{r}\frac{\partial \Psi }{\partial r}+\frac{2  \cos\theta }{r^2 \sin \theta  }\frac{\partial \Psi }{\partial \theta }=I  I'\\
-r^2\sin ^2\theta \omega \omega '\left[\left(\frac{\partial \Psi }{\partial r}\right)^2+\frac{1}{r^2}\left(\frac{\partial \Psi }{\partial \theta }\right)^2\right].
\end{split}
\end{equation}
Here $\Psi$ is the flux function; $\omega(\Psi)$ is the angular velocity of the field line and $I(\Psi)$ is the enclosed poloidal current. 

In what follows, we show one example where the membrane is spherical with a radius $r_*=1$ and rotating with angular velocity $\Omega=0.5$; the disk extends from $r_{\rm in}=2r_*$ and has an angular velocity profile $\Omega_d=r^{-3/2}$; the magnetic flux on the disk is fixed to be $\Psi(r>r_{\rm in}, \theta=\pi/2)=(r-r_0)/r_{\rm in}$, where $r_0=2r_{\rm in}$ is the location of the separatrix field line. In the steady state solution, the angular velocity of the field line is determined by the disk angular velocity at the foot point, while the current $I(\Psi)$ is given by the boundary condition (\ref{eq:membrane-boundary}). Specifically, in the slow rotation limit, we need
\begin{align}
E_{\theta}+\Omega r \sin \theta B_r=-\eta B_{\phi},\label{eq:boundary_E_theta}\\
E_{\phi}=\eta(B_{\theta}-\Omega r \sin \theta E_r).\label{eq:boundary_E_phi}
\end{align}
Since outside the membrane,
\begin{align}
B_r&=\frac{1}{r^2\sin\theta}\frac{\partial \Psi}{\partial \theta},\\
B_{\phi}&=\frac{I(\Psi)}{r\sin\theta},\\
E_{\theta}&=-\omega r\sin\theta B_r,\\
E_{\phi}&=0,
\end{align}
Equations (\ref{eq:boundary_E_theta})(\ref{eq:boundary_E_phi}) then give
\begin{align}\label{eq:boundary_GS}
I(\Psi)=-\frac{1}{\eta}(\Omega-\omega(\Psi))\sin\theta\frac{\partial\Psi}{\partial\theta}.
\end{align}

\begin{figure}
\includegraphics[width=\columnwidth]{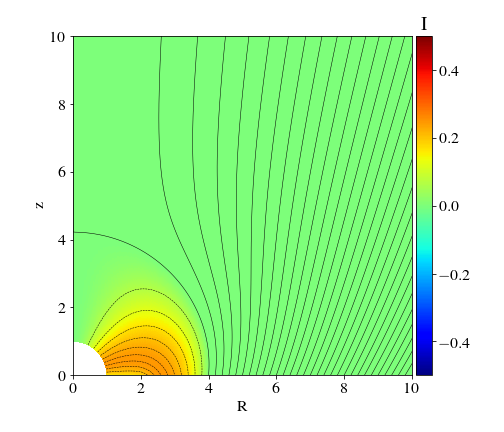}
\caption{Steady state solution for the case of a resistive membrane in flat spacetime. The membrane is spherical with a radius $r_*=1$ and rotating with angular velocity $\Omega=0.5$; the disk extends from $r_{\rm in}=2r_*$ and has an angular velocity profile $\Omega_d=r^{-3/2}$; the magnetic flux on the disk is fixed to be $\Psi(r>r_{\rm in}, \theta=\pi/2)=(r-r_0)/r_{\rm in}$, where $r_0=2r_{\rm in}$ is the location of the separatrix field line. The membrane has a resistivity $R=4\pi$. Contours show the flux function $\Psi$ and color shows the value of the enclosed poloidal current $I(\Psi)$.}\label{fig:psi_membrane}
\end{figure}

\begin{figure}
\includegraphics[width=\columnwidth]{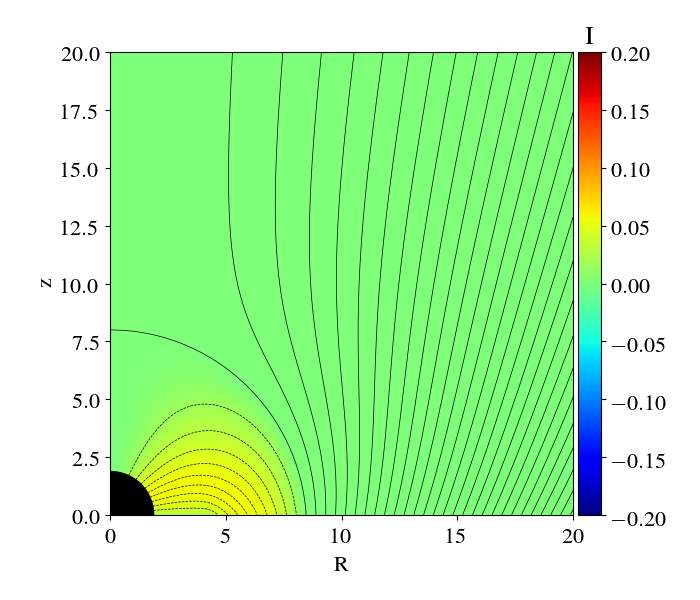}
\caption{Steady state solution for the case of a true black hole with Kerr spacetime. The black hole has a spin $a=0.5$; the disk extends outward from $r_{\rm in}=r_{\rm ISCO}$, the innermost stable circular orbit, and has an relativistic Keplerian angular velocity profile; the magnetic flux on the disk is fixed to be $\Psi(r>r_{\rm in}, \theta=\pi/2)=(r-r_0)/r_{\rm in}$, where $r_0=2r_{\rm in}$ is the location of the separatrix field line. Contours show the flux function $\Psi$ and color shows the value of the enclosed poloidal current $I(\Psi)$.}\label{fig:psi_BH}
\end{figure}

Figure \ref{fig:psi_membrane} shows an example of the steady state solution when we take the resistivity $R=4\pi$. As a comparison, Figure \ref{fig:psi_BH} shows an example of the exact solution in Kerr spacetime. Both are obtained using the method described in \citet{2019MNRAS.484.4920Y}. The qualitative similarity suggests that as far as the electromagnetic property of the black hole is concerned, it indeed behaves like a resistive membrane.

\section{The case of a dipole field}
When the current distribution in the accretion disk is only a single unidirectional loop, the magnetic field threading the central object will always open up, no matter how large the resistivity of the membrane is, how far away the current loop is from the central object, and how fast the central object is spinning (as long as $\Omega>0$). 

We can first understand this under axisymmetry constraint, using an argument similar to \citet{Uzdensky2005ApJ...620..889U}. If the field lines attached to the central object were to remain closed, the small bundle of field lines near the polar region would connect to the disk at large radius $r\to\infty$. We will show that the steady state Grad-Shafranov equation (\ref{eq:GS}) cannot be satisfied there. Firstly, these field lines have $\omega\to 0$ following their foot points on the disk (our examples in the main paper has a non-rotating disk; however, this argument is still true for a Keplerian disk).
Secondly, it is reasonable to assume that near the polar region, $\Psi\propto\theta^{\gamma}$, then boundary condition (\ref{eq:boundary_GS}) suggests that
\begin{equation}
I(\Psi)=-\frac{1}{\eta}\Omega \sin\theta\frac{\partial\Psi}{\partial\theta}\approx-\frac{1}{\eta}\Omega \gamma\Psi.
\end{equation}
So the right hand side of equation (\ref{eq:GS}) is 
\begin{equation}
rhs.=I I'\approx \frac{1}{\eta^2}\Omega^2\gamma^2\Psi.
\end{equation}
Now the left hand side of equation (\ref{eq:GS}), without the $\omega$ term, is a diffusion-like operator and roughly scales as $\Psi/r^2$. As a result, near the foot points of the polar field lines on the disk, the left hand side is $\sim1/r^2$ times the right hand side and the equation cannot be satisfied for very large $r$. We can also see that for the equilibrium to exist, we need the foot point distance to not be much larger than $r\sim\eta/\Omega$.

\begin{figure}
\includegraphics[width=\columnwidth]{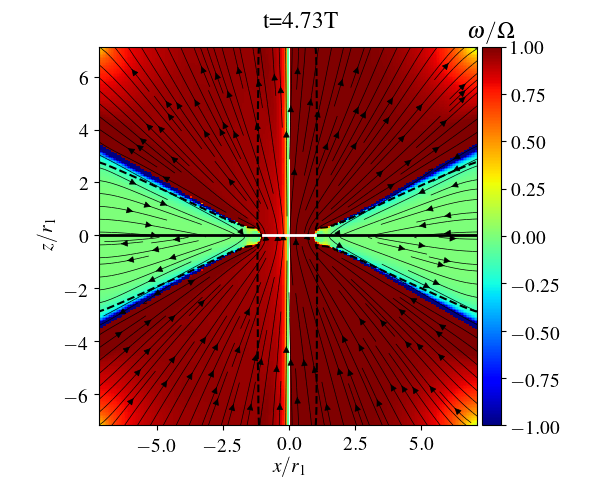}
\caption{Final steady state for the case where the central membrane has zero resistivity and the initial current distribution is a single unidirectional loop located at $\sim 1.08r_1$. The angular velocity of the membrane is such that $\Omega r_1/c=0.9$. In the plot, the streamlines show the poloidal magnetic field, and color shows the angular velocity of the field line. The dashed lines correspond to the light surfaces.}\label{fig:dipole_rin2_54_eta0}
\end{figure}

\begin{figure}
\includegraphics[width=\columnwidth]{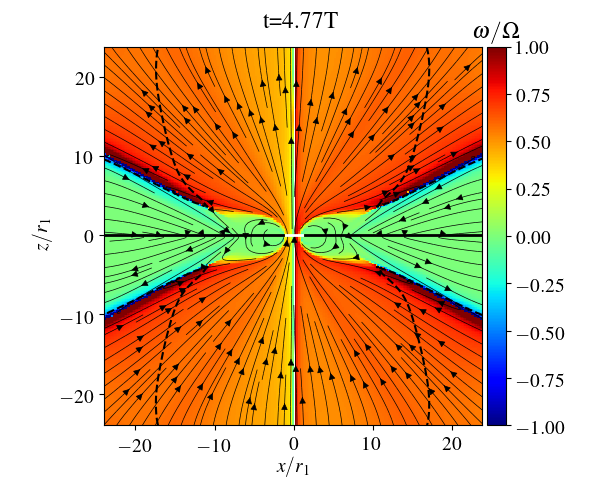}
\caption{Final steady state for the case where the central membrane has resistivity $\eta=1$ and the initial current distribution is a single unidirectional loop located at $\sim 1.4r_1$. The angular velocity of the membrane is such that $\Omega r_1/c=0.1$. In the plot, the streamlines show the poloidal magnetic field, and color shows the angular velocity of the field line. The dashed lines correspond to the light surfaces.}\label{fig:dipole_rin2_70_eta1}
\end{figure}

In our time-dependent force-free simulations, we indeed get steady, axisymmetric final solutions for these cases. Figures \ref{fig:dipole_rin2_54_eta0} and \ref{fig:dipole_rin2_70_eta1} show two examples. As the dipole field opens up, there exists a current sheet separating the outgoing and ingoing magnetic fluxes; the field attached to the central object is rotating while that attached to the accretion disk is not rotating (as we set $\Omega_d=0$). At large distances, the current sheet lies along a polar angle around $\theta=60^{\circ}$. We can understand this opening angle as follows. Since the initial current distribution is isolated, the rotation induced outgoing flux solution is asymptotically split-monopole-like. 
Due to magnetic flux conservation, the amount of outgoing magnetic flux should be equal to the ingoing flux. Therefore, the separation lies at $\theta\approx60^{\circ}$.



\bsp	
\label{lastpage}
\end{document}